\documentclass[
  aps,
  prl,
  reprint,
  superscriptaddress,
  amsfonts,amssymb,amsmath
]{revtex4-2}

\usepackage{newtxtext}
\usepackage{bm}
\usepackage{latexsym}
\usepackage{xcolor}
\usepackage{physics}
\usepackage{hyperref}
\hypersetup{
  setpagesize  = false,
  colorlinks   = true,
  urlcolor     = blue,
  linkcolor    = blue,
  citecolor    = blue
}
\usepackage{tikz}

\newcommand{\beginendmatt}{
  \twocolumngrid
  \setcounter{equation}{0}
  \renewcommand{\theequation}{A\arabic{equation}}%
  \setcounter{figure}{0}
  \renewcommand{\thefigure}{A\arabic{figure}}%
  \setcounter{table}{0}
  \renewcommand{\thetable}{A\arabic{table}}%
  \setcounter{section}{0}
  \renewcommand{\thesection}{A\Roman{section}}%
  \makeatletter
}

\newcommand{\beginsupplement}{
  \setcounter{equation}{0}
  \renewcommand{\theequation}{S\arabic{equation}}%
  \setcounter{figure}{0}
  \renewcommand{\thefigure}{S\arabic{figure}}%
  \setcounter{table}{0}
  \renewcommand{\thetable}{S\arabic{table}}%
  \setcounter{section}{0}
  \renewcommand{\thesection}{S\Roman{section}}%
  \makeatletter
}

\begin{document}

\title{Stabilizer R\'{e}nyi Entropy Encodes Fusion Rules of Topological Defects and Boundaries}

\author{Masahiro Hoshino}
\email{hoshino-masahiro921@g.ecc.u-tokyo.ac.jp}
\affiliation{Department of Physics, The University of Tokyo, 7-3-1 Hongo, Bunkyo-ku, Tokyo 113-0033, Japan}
\author{Yuto Ashida}
\affiliation{Department of Physics, The University of Tokyo, 7-3-1 Hongo, Bunkyo-ku, Tokyo 113-0033, Japan}
\affiliation{Institute for Physics of Intelligence, The University of Tokyo, 7-3-1 Hongo, Bunkyo-ku, Tokyo 113-0033, Japan}

\date{\today}

\begin{abstract}
  We demonstrate that the stabilizer R\'{e}nyi entropy (SRE), a computable measure of quantum magic, can serve as an information-theoretic probe for universal properties associated with conformal defects in one-dimensional quantum critical systems. Using boundary conformal field theory, we show that open boundaries manifest as a universal logarithmic correction to the SRE, whereas topological defects yield a universal size-independent term. When multiple defects are present, we find that the universal terms in the SRE faithfully reflect the defect-fusion rules that define a noninvertible symmetry algebra. These analytical predictions are corroborated by numerical calculations of the Ising model, where boundaries and topological defects are described by Cardy states and Verlinde lines, respectively.
\end{abstract}

\maketitle

Understanding universal properties of quantum resources is a cornerstone of modern physics.
A paradigmatic example is entanglement, whose universal behavior has provided profound insights into various fields ranging from condensed matter to high-energy physics~\cite{amico2008entanglement,zeng2019quantum}.
In one-dimensional (1D) quantum critical systems described by conformal field theory (CFT)~\cite{difrancesco1997conformal}, for instance, the entanglement entropy (EE) exhibits a universal logarithmic scaling governed by the central charge~\cite{holzhey1994geometric,vidal2003entanglement,calabrese2004entanglement,calabrese2009entanglement}.
This seminal result has been extended to systems with conformal defects, revealing an even richer tapestry of universal features~\cite{brehm2015entanglement,gutperle2016note,jiang2017entanglement,cornfeld2017entanglement,roy2022entanglement,roy2022entanglementa,rogerson2022entanglement,roy2024topological,gutperle2024note}.

Beyond entanglement, a crucial resource for universal quantum computation is nonstabilizerness (also known as quantum magic)~\cite{bravyi2005universal,veitch2014resource,chitambar2019quantum}.
The Gottesman-Knill theorem~\cite{gottesman1998heisenberg,nielsen2010quantum} establishes that quantum circuits composed solely of Clifford gates can be efficiently simulated on a classical computer; thus, non-Clifford operations constitute the essential ingredient for quantum computational advantage.
The stabilizer R\'{e}nyi entropy (SRE) has recently emerged as a computable measure of nonstabilizerness~\cite{leone2022stabilizer,leone2024stabilizer}, enabling a surge of investigations into its role in quantum phase transitions~\cite{oliviero2022magicstate,tarabunga2023manybody,frau2024nonstabilizerness,falcao2025nonstabilizerness,viscardi2025interplay}, many-body dynamics and thermalization~\cite{rattacaso2023stabilizer,tirrito2024anticoncentration,szombathy2025independent,odavic2025stabilizer,szombathy2025spectral,turkeshi2025pauli,turkeshi2025magic,sticlet2025nonstabilizerness,tirrito2025universal}, measurement-induced phenomena~\cite{tarabunga2024magic,niroula2024phase,fux2024entanglement}, numerical methods~\cite{haug2023quantifying,haug2023stabilizer,lami2023nonstabilizerness,tirrito2024quantifying,tarabunga2024nonstabilizernessa,passarelli2024nonstabilizerness,liu2024nonequilibrium,collura2024quantum,ding2025evaluating,russomanno2025efficient,kozic2025computing,tarabunga2025efficient}, and other related topics~\cite{odavic2023complexity,turkeshi2023measuring,cao2025gravitational,bera2025nonstabilizerness,jasser2025stabilizer,sinibaldi2025nonstabilizerness,smith2025nonstabilizerness}.
In 1D critical systems, it has been shown that the SRE acquires a universal size-independent term governed by the Affleck-Ludwig $g$-factor~\cite{affleck1991universal}, which encapsulates universal data of the underlying CFT~\cite{hoshino2025stabilizer}.

\begin{figure}[b]
  \centering
  \includegraphics[width=\linewidth, clip]{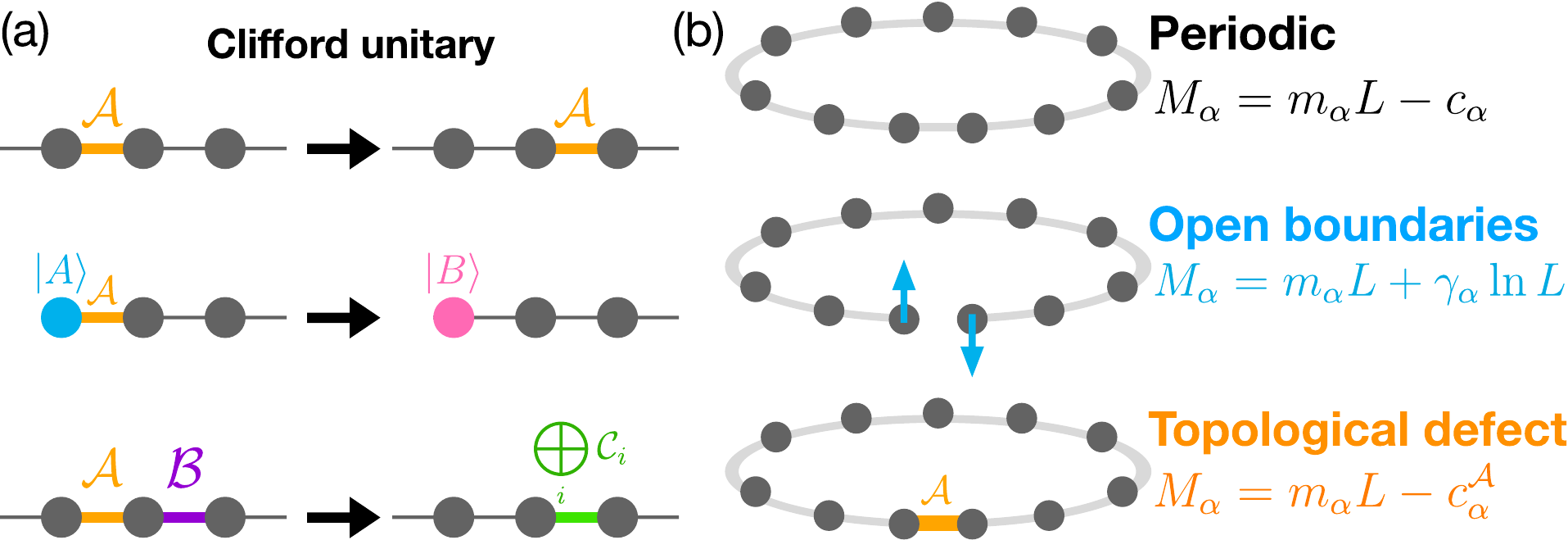}
  \caption{(a)~Conformal defects can be moved and fused by Clifford unitaries. The stabilizer R\'{e}nyi entropy remains invariant under such operations, allowing it to directly probe the algebraic structure of defect fusion. (b)~Open boundaries can be viewed as factorizing defects inserted in a periodic chain, whereas topological defects are created by locally altering the Hamiltonian.}
  \label{fig:conceptart}
\end{figure}

Concurrently, the physics of conformal defects has garnered renewed interest in the context of noninvertible symmetries, where the concept of global symmetry has been generalized beyond the traditional group-theoretic framework~\cite{frohlich2004kramerswannier,frohlich2007duality,davydov2012invertible}.
In quantum field theories, generalized symmetries are embodied by topological defects, i.e., operators supported on extended manifolds that can be freely deformed without affecting physical observables~\cite{gaiotto2015generalized}.
The algebraic structure of these symmetries is captured by fusion rules of topological defects~\cite{bhardwaj2018finite,chang2019topological,thorngren2024fusiona,thorngren2024fusion}.
There, the fusion of noninvertible defects can yield a superposition of multiple defect channels, unlike their invertible counterparts.
This richer algebraic structure potentially imposes unforeseen constraints on quantum resources in many-body systems, motivating the search for an information-theoretic probe of their manifestations.

The aim of this Letter is to analyze and propose the SRE as a natural probe for universal properties associated with conformal defects.
Our central finding stems from a defining property of the SRE, namely, its invariance under Clifford unitaries.
Specifically, we demonstrate that Clifford operations implement movement and fusion of topological defects in lattice models (Fig.~\ref{fig:conceptart}(a)).
Consequently, the SRE remains invariant when conformal defects are moved/fused, in contrast to other information quantities like EE, which are sensitive to defect locations and the choice of subsystems.
For instance, in the case of the Ising CFT, EE fails to distinguish between the identity defect and the $\mathbb{Z}_2$ defect~\cite{roy2022entanglement,roy2022entanglementa}, whereas the SRE can discern their distinct universal contributions.
The SRE thus serves not merely as a probe but as a discovery tool; namely, through systematically searching for lattice operators that yield universal scaling behaviors, one can identify lattice realizations of topological defects and their fusion rules without prior knowledge.

Using boundary CFT, we show that factorizing defects, or equivalently, open boundaries induce a universal logarithmic correction to the SRE, whereas the influence of topological defects is captured by a universal size-independent term (Fig.~\ref{fig:conceptart}(b)).
Crucially, when multiple defects are present, the universal terms are governed by the underlying fusion rules.
We corroborate these field-theoretical results with tensor-network calculations of the Ising model.
Our work thus puts forward the SRE not only as a measure of a resource but as an information-theoretic probe for the algebraic structure of generalized symmetries.

From a broader perspective, our results provide insights into advancing tensor-network methods, such as the recently proposed Clifford augmented matrix product states~\cite{lami2024quantum,fux2024disentangling,qian2024augmenting,masot-llima2024stabilizer,qian2025clifford,mello2025clifforddressed,nakhl2025stabilizer,fu2025clifford}.
For instance, we show that the Clifford unitaries used to disentangle critical spin chains~\cite{frau2025stabilizer,fan2025disentangling} can be understood as operations that move and fuse a topological defect with a boundary.
Furthermore, given the growing experimental capability to manipulate defects and to measure magic-related quantities in quantum platforms~\cite{bluvstein2024logical,monroe2021programmable}, our findings might motivate an experimental exploration for the physics of generalized symmetries.

\textit{Many-body quantum magic with conformal defects.---}
The constraints imposed by conformal invariance on line defects in a $(1{+}1)$D CFT have been studied both within continuum~\cite{graham2004defect,quella2007reflection,petkova2013topological,buican2017anyonic,bachas2013fusion,lin2023duality,chang2023topological,choi2023remarks,choi2024selfduality,furuta2024classification} and lattice~\cite{oshikawa1997boundary,aasen2016topological,hauru2016topological,belletete2020topological,aasen2020topological,fukusumi2021open,ashkenazi2022duality,lootens2023dualities,belletete2023topological,li2023noninvertible,fukusumi2024protected,seifnashri2024liebschultzmattis,khan2024quantum,seiberg2024noninvertible,mana2024kennedytasaki,sinha2024lattice,okada2024noninvertible,seifnashri2025cluster,pace2025lattice,li2025intrinsically,zhang2025kramerswannier}.
A general conformal defect serves as an interface between two theories, $\text{CFT}_1$ and $\text{CFT}_2$, and is defined by the conservation of the energy-momentum tensor across it, i.e., $T_1\,{-}\,\overline{T}_1 \,{=}\, T_2\,{-}\,\overline{T}_2$, where $T_i$ and $\overline{T}_i$ are the holomorphic and antiholomorphic components of the energy-momentum tensor for $\text{CFT}_i$, respectively.
Open boundaries are totally reflective, meaning the conservation law holds for each theory independently $T_i \,{-}\, \overline{T}_i \,{=}\, 0$.
In contrast, topological defects are totally transmissive, satisfying $T_1 \,{=}\, T_2$ and $\overline{T}_1 \,{=}\, \overline{T}_2$.
On a lattice, such conformal defects can be realized by locally modifying the Hamiltonian as illustrated in Fig.~\ref{fig:conceptart}(b).
In diagonal Virasoro minimal models, such as the Ising CFT, open boundaries are classified by Cardy states~\cite{cardy1989boundary} and topological defects by Verlinde lines~\cite{verlinde1988fusion,petkova2001generalised}.

We consider the ground state $\ket*{\psi}$ of a critical spin chain of $L$ qubits with conformal defects and measure its nonstabilizerness using the SRE defined as follows:
\begin{equation}\label{eq:sre}
  M_{\alpha}(\psi) = \frac{1}{1-\alpha}\ln\sum_{\vec{m}}\frac{\Tr^{2\alpha}[\sigma^{\vec{m}}\psi]}{2^L}.
\end{equation}
Here, $\psi \,{=}\, \ketbra*{\psi}{\psi}$ represents the density matrix corresponding to $\ket*{\psi}$, and the $L$-qubit Pauli string $\sigma^{\vec{m}}$ is given by
\begin{equation}\label{eq:pauli-string}
  \sigma^{\vec{m}} = \bigotimes_{j=1}^{L} \sigma^{m_{2j{-}1}m_{2j}}\quad (\vec{m}\in\qty{0,1}^{2L}),
\end{equation}
where the Pauli matrices are labeled as $\sigma^{00}{=}\,I,\sigma^{10}{=}\,X,\sigma^{01}{=}\,Z$, and $\sigma^{11}{=}\,Y$.
The SRE has the following key properties: (i) faithfulness, $M_{\alpha}(\psi) \,{=}\, 0$ if and only if $\ket*{\psi}$ is a stabilizer state, (ii) stability under Clifford unitaries $C\,{\in}\,C_L$, $M_{\alpha}(C\psi C^\dag) \,{=}\, M_{\alpha}(\psi)$, and (iii) additivity, $M_{\alpha}(\psi\,{\otimes}\,\phi) \,{=}\, M_{\alpha}(\psi)\,{+}\,M_{\alpha}(\phi)$.
SREs with $\alpha\,{\geq}\,2$ also satisfy monotonicity under stabilizer protocols~\cite{leone2024stabilizer}.
We recall that the Clifford group $C_L$ consists of unitary transformations that preserve the Pauli group, i.e., $C\sigma^{\vec{m}} C^\dag \,{=}\, \sigma^{\vec{m}'}$ for $C\,{\in}\, C_L$; stabilizer states are constructed by applying Clifford unitaries to the product state $\ket*{0}^{\otimes L}$.

\begin{figure}[tb]
  \centering
  \includegraphics[width=\linewidth, clip]{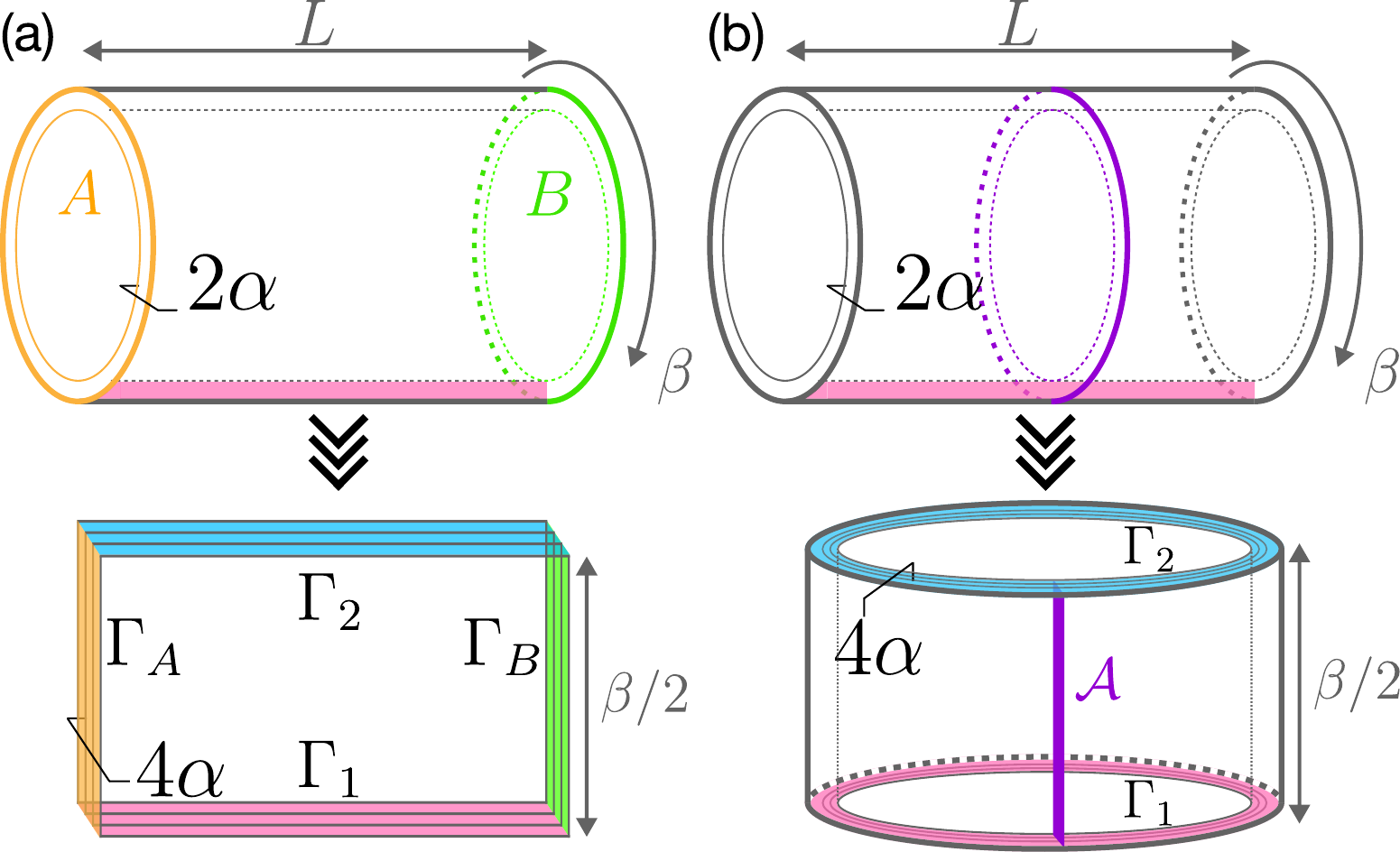}
  \caption{(a)~When the system has open boundaries, the partition function $Z_{2\alpha}$ of the $2\alpha$-component theory in Eq.~\eqref{eq:replica-trick} is defined on the cylinder with two ends $A,B$ and a horizontal line defect due to the Bell projection at the fixed imaginary time $\tau\,{=}\,0$. After folding, the cylinder becomes a rectangle with four boundaries, and the number of fields is double. (b)~Torus of size $L\,{\times}\,\beta$ with a topological defect $\mathcal{A}$ can be folded into a cylinder of circumference $L$ and length $\beta/2$ with a vertical line defect. Periodicity in the spatial direction is implicit in the top panel.}
  \label{fig:partition-function}
\end{figure}

Our CFT analysis starts by expressing the SRE as the participation entropy of a doubled state in the Bell basis~\cite{hoshino2025stabilizer}.
Using the Choi-Jamiolkowski isomorphism~\cite{jamiolkowski1972linear,choi1975completely}, we obtain
\begin{equation}\label{eq:sre-participationE}
  M_{\alpha}(\psi) = \frac{1}{1-\alpha}\ln\sum_{\vec{m}} \Tr^{\alpha}[P^{\vec{m}}(\psi\otimes\psi^\ast)] - (\ln 2)L.
\end{equation}
Here, $P^{\vec{m}}{=}\bigotimes_{j=1}^{L}\ketbra*{\mathrm{Bell}^{m_{2j{-}1}m_{2j}}}{\mathrm{Bell}^{m_{2j{-}1}m_{2j}}}$ is the projection operator onto the Bell basis defined as $\ket*{\mathrm{Bell}^{b_1b_2}} \,{=}\, (\ket*{0b_1}{+}(-1)^{b_2}\ket*{1\overline{b}_1})/\sqrt{2}$ with $b_1b_2\,{\in}\,\qty{0,1}^{2}$.
Thus, the SRE can be interpreted as the R\'{e}nyi entropy of the classical probability distribution $p_{\vec{m}}=\Tr[P^{\vec{m}}(\psi\otimes\psi^\ast)]$, up to a constant offset.
In the path-integral formalism, a projection operator manifests as the Euclidean boundary action with an infinitely large coupling~\cite{garratt2023measurements,sun2023new,weinstein2023nonlocality,lee2023quantum,yang2023entanglement,murciano2023measurementaltered,ashida2024systemenvironment,zhang2025universal,hoshino2025entanglement,popov2025factorizing,naus2025practical}, which enforces the field configuration that can lead to a conformal defect along the \textit{spatial} direction.
Using the replica trick for $\alpha\in\mathbb{Z}_{\geq 2}$, we thus obtain
\begin{align}
  \label{eq:sre-partition-function}
  M_{\alpha}(\psi)
   & = \frac{1}{1-\alpha}\ln\left(\frac{Z_{2\alpha}}{Z^{2\alpha}}\right){-}(\ln 2)L,               \\
  \label{eq:replica-trick}
  \frac{Z_{2\alpha}}{Z^{2\alpha}}
   & := \Tr[\sum_{\vec{m}}(P^{\vec{m}})^{\otimes \alpha} (\psi\otimes\psi^\ast)^{\otimes \alpha}].
\end{align}
Here, $Z_{2\alpha}$ is a partition function of the $2\alpha$-component CFT where the projection $\sum_{\vec{m}}(P^{\vec{m}})^{\otimes \alpha}$ yields a line defect at the imaginary-time slice (horizontal lines in top panels of Fig.~\ref{fig:partition-function}), and $Z$ is the original partition function for normalization.

In contrast, a defect contained in the Hamiltonian is described by a line defect along the \textit{imaginary-time} direction.
When this defect is factorizing, the geometry of the partition function $Z_{2\alpha}$ is a cylinder of length $L$ and circumference $\beta$ with two boundaries $A$ and $B$ at both ends (top panel of Fig.~\ref{fig:partition-function}(a)); here, the inverse temperature $\beta$ is understood to be taken infinitely large for the ground state.
We use the folding trick to rewrite $Z_{2\alpha}$ as a partition function of $4\alpha$ components on an $L\times \beta/2$ rectangle with the following four boundary conditions: $\Gamma_1$ at $\tau\,{=}\,0$ subject to the projection $\sum_{\vec{m}}(P^{\vec{m}})^{\otimes \alpha}$, $\Gamma_2$ at $\tau\,{=}\,\beta/2$ corresponding to the artificial boundary created by the folding, and $\Gamma_{A,B}$ being the doubled boundary of the original one at each end $A,B$ (bottom panel of Fig.~\ref{fig:partition-function}(a)).
Meanwhile, when the Hamiltonian contains a topological defect, the folding reduces $Z_{2\alpha}$ to a partition function of $4\alpha$ components on a cylinder of circumference $L$ and length $\beta/2$.
The resulting cylinder has the boundaries $\Gamma_1$ and $\Gamma_2$ at the ends $\tau=0,\beta/2$, while the topological defect denoted by $\mathcal{A}$ is inserted along the imaginary-time direction (Fig.~\ref{fig:partition-function}(b)).

One of our main results is that the SRE captures universal properties of conformal defects in qualitatively different ways, depending on the nature of defects.
With factorizing defects, the SRE acquires the universal logarithmic correction due to the conical singularity at the corners~\cite{cardy1988finitesize,kleban1991free,imamura2006boundary,zaletel2011logarithmic,stephan2014shannon}.
Specifically, in the zero-temperature limit $\beta\,{\to}\,\infty$, we obtain
\begin{equation}\label{eq:free-energy-factorizing}
  -\ln\frac{Z_{2\alpha}}{Z^{2\alpha}} = b_{\alpha} L + \gamma_{\alpha} \ln L + O(1),
\end{equation}
where $b_{\alpha}$ is a nonuniversal line energy density associated with $\Gamma_1$.
The universal quantity $\gamma_{\alpha}$ is the sum of the contributions from each corner of the spacetime manifold.
In general, for a corner of angle $\theta$ separating boundaries $a$ and $b$, such contribution is given by~\cite{cardy1988finitesize}
\begin{equation}\label{eq:conical-singularity}
  \gamma^{ab} := \frac{c}{24}\qty(\frac{\theta}{\pi} - \frac{\pi}{\theta}) + \frac{\pi}{\theta}h_{ab},
\end{equation}
where $c$ is the central charge, and $h_{ab}$ is the scaling dimension of the corresponding boundary condition changing operator (BCCO)~\cite{affleck1997boundary,recknagel2013boundary}.
In the present geometry (Fig.~\ref{fig:partition-function}(a)), the relevant corners are formed by the intersection of the spatial boundary $\Gamma_{A/B}$ and the measurement-induced boundary $\Gamma_1$ at $\tau\,{=}\,0$ with the angle $\theta\,{=}\,\pi/2$, and we replace $c\,{\to}\,4\alpha c$ because of the $4\alpha$ components after folding.
The contributions associated with the artificial boundary $\Gamma_2$ vanish by construction, while the two corners at the intersections of $\Gamma_1$ and $\Gamma_{A/B}$ contribute, yielding $\gamma_{\alpha}=\gamma^{1A}+\gamma^{1B}$.

In contrast, when topological defects are present, the logarithmic correction is absent and, instead, the universal data are encoded in a size-independent term:
\begin{equation}\label{eq:free-energy-topological}
  -\ln\frac{Z_{2\alpha}}{Z^{2\alpha}} = b_{\alpha}L -\ln g_{\alpha}^\mathcal{A} + o(1),
\end{equation}
where $g_{\alpha}^\mathcal{A} = \braket*{0_\mathcal{A}}{\Gamma_1}$ is the $g$-factor given by the overlap between the boundary state $\ket*{\Gamma_1}$ and the ground state $\ket*{0_\mathcal{A}}$ of the sector containing the topological defect $\mathcal{A}$ (bottom panel of Fig.~\ref{fig:partition-function}(b)).
Here, the correction term $o(1)$ arises solely from finite-size effects (e.g., $O(L^{-1})$ corrections) and vanishes in the thermodynamic limit.
This is in contrast to the nonuniversal $O(1)$ term in Eq.~\eqref{eq:free-energy-factorizing}.
Altogether, the SRE isolates the universal defect properties through distinct features: a logarithmic correction with coefficient $(\alpha\,{-}\,1)^{-1}\gamma_{\alpha}$ for factorizing defects, and a size-independent term $(1\,{-}\,\alpha)^{-1}\ln g_{\alpha}^\mathcal{A}$ for topological defects (cf. Eq.~\eqref{eq:sre-partition-function}).

\begin{figure}[b]
  \centering
  \includegraphics[width=\linewidth, clip]{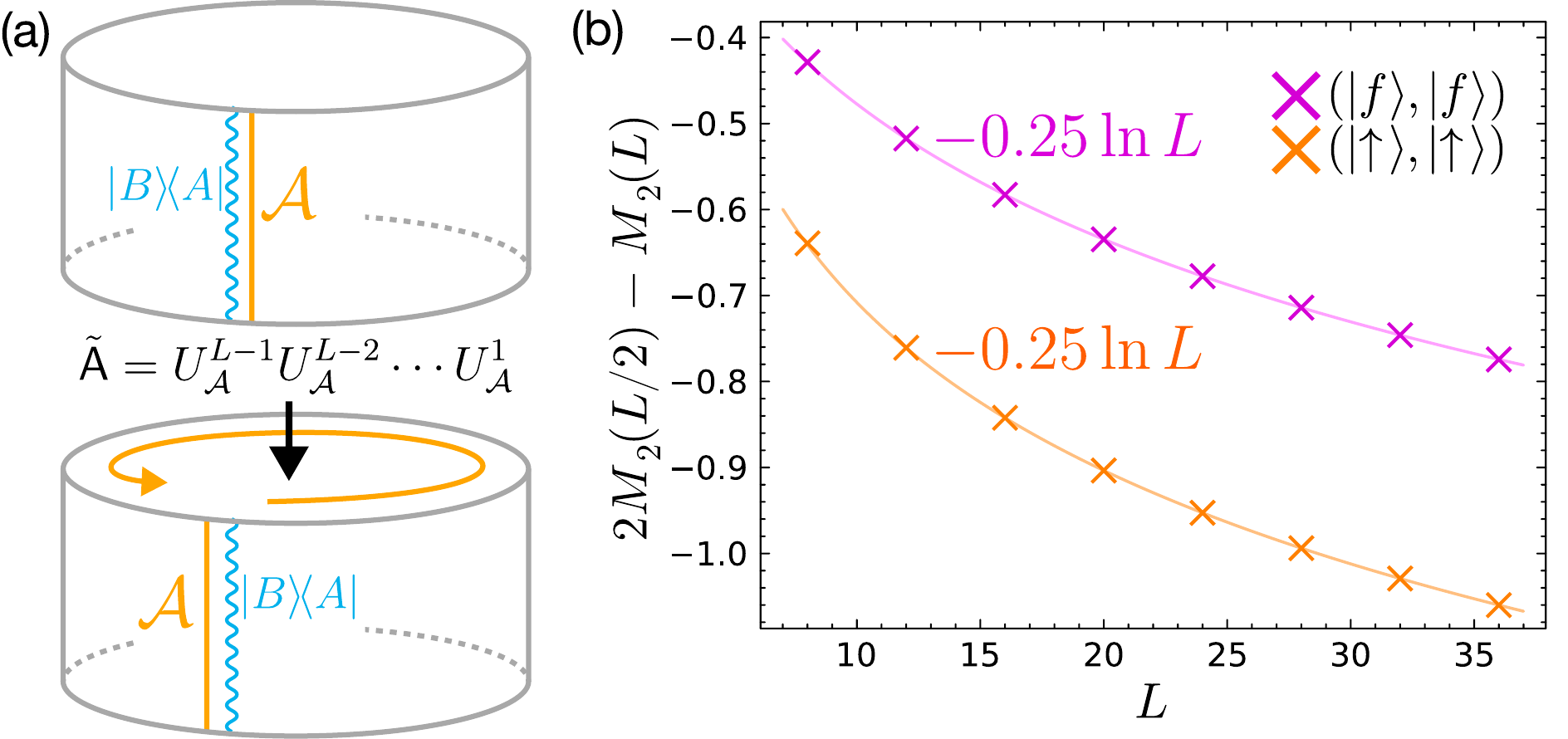}
  \caption{(a)~The SREs of open chains with the boundary pairs $(\mathcal{A}\,{\ast}\ket*{A},\ket*{B})$ and $(\ket*{A},\mathcal{A}\,{\ast}\ket*{B})$ are equivalent provided that one can move the defect $\mathcal{A}$ from one side to the other and fuse it with the boundary via a Clifford unitary.
    (b)~The universal logarithmic correction is extracted by fitting the data $2M_2(L/2)-M_2(L)$ as a function of the system size $L$, where $M_2(L)$ is the $\alpha=2$ SRE of the $L$-qubit Ising critical state. The estimated coefficient agrees with the theoretical value $-1/4$ in both the $(\ket*{f},\ket*{f})$ and $(\ket*{\uparrow},\ket*{\uparrow})$ boundary pairs.}
  \label{fig:sre_factorizing}
\end{figure}

Another key observation is that the SRE can probe dynamical properties such as fusion rules of conformal defects.
This unique capability arises from the invariance of the SRE under Clifford unitaries.
Indeed, the known transformations that move and fuse topological defects are implemented by Clifford unitaries~\cite{fukusumi2021open,seiberg2024noninvertible,pace2025lattice}.
Given that defect operations naturally map a Pauli string to another Pauli string, a defining feature of Clifford unitaries, we posit that the movement and fusion of a topological defect $\mathcal{A}$ can be realized by Clifford unitaries.

This proposition leads to two primary consequences in the universal features of the SRE:
\begin{itemize}
  \item[(I)] The universal coefficient of the logarithmic term is unchanged while a defect $\mathcal{A}$ is moved and fused with either boundary; for instance, the coefficient for the pair of boundaries $(\mathcal{A}\,{\ast}\ket*{A},\ket*{B})$ is equivalent to that for $(\ket*{A},\mathcal{A}\,{\ast}\ket*{B})$ as illustrated in Fig.~\ref{fig:sre_factorizing}(a).
  \item[(II)] When the Hamiltonian contains multiple topological defects, the universal size-independent term is determined by the lowest-energy sector appearing in their fusion rule.
\end{itemize}
In property (I), we denote the fusion of a topological defect $\mathcal{A}$ with a boundary $A$ as $\mathcal{A}\,{\ast}\ket*{A}$.
Property (II) reflects the fact that a fusion rule $\mathcal{A}\,{\otimes}\,\mathcal{B}=\bigoplus_{i}\mathcal{C}_i$ indicates that a Hamiltonian $H_{\mathcal{A}{\otimes}\mathcal{B}}$ with two topological defects can be mapped by Clifford unitaries to a direct sum of single-defect Hamiltonians $\bigoplus_i H_{\mathcal{C}_i}$.
Consequently, the lowest-energy sector gives the ground state of the original system, and the universal term in the SRE is governed by the $g$-factor of this defect channel.
In what follows, we demonstrate these properties through a concrete analysis of the Ising CFT.

\textit{Factorizing defects in the Ising CFT.---}
A standard lattice model that realizes the $c=1/2$ Ising CFT is the transverse field Ising model,
\begin{equation}\label{eq:tfim}
  H = -\sum_{j=1}^{L}(Z_{j}Z_{j+1} + \lambda X_j),
\end{equation}
where we impose periodic boundary conditions $X_j\,{=}\,X_{j{+}L}$ and $Z_j\,{=}\,Z_{j{+}L}$, $\lambda$ represents the transverse field, and $\lambda=1$ is the critical point.
There are three conformal boundary states in the Ising CFT~\cite{cardy1989boundary}: the free state $\ket*{f}$, and the spin-up $\ket*{\uparrow}$ and -down $\ket*{\downarrow}$ states.
Open boundaries, or equivalently, factorizing defects can be created by locally modifying the periodic chain.
For example, a factorizing defect with free boundaries at both ends can be realized by setting $Z_{L}Z_{1}\mapsto0$, while a defect with spin-up boundaries at both ends corresponds to $Z_{L}Z_{1}\mapsto -Z_{L}-Z_{1}$ in Eq.~\eqref{eq:tfim}.

As inferred from Eq.~\eqref{eq:conical-singularity}, the field-theoretical derivation of the universal coefficient $\gamma_\alpha=\gamma^{1A}\,{+}\,\gamma^{1B}$ boils down to computing the scaling dimensions $h_{1A/B}$ of the BCCOs between $\Gamma_1$ and $\Gamma_{A/B}$ in the $4\alpha$-component replicated CFT.
To this end, we map the doubled Ising CFT to a single $S^1/\mathbb{Z}_2$ CFT, which is the $\mathbb{Z}_2$ orbifold of the free-boson CFT compactified as $\phi\sim\phi+2\pi$~\cite{difrancesco1987critical,ginsparg1988curiosities}.
This mapping to the free-boson CFT makes the analysis of multicomponent theories more tractable.
Under this mapping, the pair of free boundary states $\ket*{ff}$ becomes the Dirichlet boundary state $\ket*{D(\pi/2)}_{\mathrm{orb}}$ localized at $\phi=\pi/2$~\cite{oshikawa1997boundary}.
The corresponding boundary state in the replicated theory $Z_{2\alpha}$ is $\ket*{\Gamma_f}=\ket*{D(\pi/2)}_{\mathrm{orb}}^{\otimes 2\alpha}$.
Since $\Gamma_1$ corresponds to a mixed Dirichlet-Neumann boundary with a single component obeying the Neumann boundary condition~\cite{hoshino2025stabilizer}, the scaling dimension of the BCCO between $\Gamma_1$ and $\Gamma_f$ equals that of a twist (spin) field $\sigma$, which acts as changing a single boson field boundary from Dirichlet to Neumann.
This allows us to determine $h_{1f}=1/16$, which can also be confirmed by explicitly evaluating the amplitude $\mel*{\Gamma_f}{e^{-\frac{\beta}{2} H_{\text{CFT}}}}{\Gamma_1}$, where $H_{\text{CFT}}$ is the CFT Hamiltonian on a ring (see Supplemental Material~\cite{SM} for details).

Interestingly, property (I) above allows us to immediately determine other scaling dimensions such as $h_{1\uparrow}$ and $h_{1\downarrow}$.
To see this, we recall that the Ising CFT has three topological defects: the identity $1$, the invertible $\mathbb{Z}_2$ defect $\eta$, and the noninvertible duality defect $\mathcal{D}$.
The $\eta$ and $\mathcal{D}$ defects act on the boundary states as follows:
\begin{align}
  \eta\,{\ast}\ket*{f}        & = \ket*{f},                            & \eta\,{\ast}\ket*{\uparrow}        & = \ket*{\downarrow}, & \eta\,{\ast}\ket*{\downarrow}        & = \ket*{\uparrow}, \label{eq:fusion-boundary-eta} \\
  \mathcal{D}\,{\ast}\ket*{f} & = \ket*{\uparrow}{+}\ket*{\downarrow}, & \mathcal{D}\,{\ast}\ket*{\uparrow} & = \ket*{f},          & \mathcal{D}\,{\ast}\ket*{\downarrow} & = \ket*{f}.\label{eq:fusion-boundary-d}
\end{align}
These relations are consistent with the following fusion rules among the topological defects:
\begin{equation}
  \eta\otimes\eta = 1,\quad      \eta\otimes\mathcal{D} = \mathcal{D}\otimes\eta = \mathcal{D},\quad \mathcal{D}\otimes\mathcal{D}  = 1\oplus\eta.
\end{equation}
The fusion rule~\eqref{eq:fusion-boundary-eta} and property (I) of the SRE allows us to conclude that the SRE with the pair of boundaries $(\ket*{\uparrow},\ket*{f})$ is equivalent to that with the pair $(\ket*{\downarrow},\ket*{f})$, implying $h_{1\uparrow}=h_{1\downarrow}$.
Moreover, the relation~\eqref{eq:fusion-boundary-d} together with the property (I) indicates that the pair $(\ket*{f},\ket*{f})$ must give the same SRE as the pair $(\ket*{\uparrow},\ket*{\uparrow}{+}\ket*{\downarrow})$.
Since we have $h_{1\uparrow}=h_{1\downarrow}$, it follows that $h_{1\uparrow}=h_{1(\uparrow{+}\downarrow)}$.
Consequently, we get $h_{1\uparrow}=h_{1\downarrow}=h_{1f}=1/16$, indicating that, for all nine elementary factorizing defects, the SRE should behave as
\begin{equation}\label{eq:sre-open}
  M_{\alpha}(\psi) = m_{\alpha}L - \frac{1}{4}\ln L + O(1),
\end{equation}
where we used Eq.~\eqref{eq:conical-singularity} to obtain $\gamma^{1A/B}=(1-\alpha)/8$ with $c=2\alpha$, $\theta=\pi/2$, and $h_{1A/B}=1/16$.
This result is numerically verified by calculating the SRE of open chains using the replica-Pauli MPS method~\cite{tarabunga2024nonstabilizernessa} (Fig.~\ref{fig:sre_factorizing}(b)).

It is noteworthy that local movement of the defects $\eta$ and $\mathcal{D}$ can be realized by $U_{\eta}^j=X_j$ and $U_{\mathcal{D}}^j=\mathsf{CZ}_{j,j{+}1}\mathsf{H}_j$, respectively, where $\mathsf{H}_j$ is the Hadamard gate and $\mathsf{CZ}_{j,k}$ is the CZ gate, both being Clifford gates.
Similarly, the fusion of defects can also be implemented by Clifford unitaries (see End Matter).
Meanwhile, the global movement $\tilde{\mathsf{D}}=U_{\mathcal{D}}^{L-1}U_{\mathcal{D}}^{L-2}\cdots U_{\mathcal{D}}^1$ of the noninvertible duality defect $\mathcal{D}$ satisfies $\tilde{\mathsf{D}} X_j \,{=}\, Z_{j{-}1}Z_j\tilde{\mathsf{D}}$ and $\tilde{\mathsf{D}}Z_jZ_{j{+}1} \,{=}\, X_j\tilde{\mathsf{D}}$ except at the ends of the chain.
This observation gives a simple theoretical explanation of the disentangling Clifford unitary numerically found in Ref.~\cite{fan2025disentangling}; it is nothing but the defect-movement operation $\tilde{\mathsf{D}}$, which acts as the Kramers-Wannier transformation in the bulk while altering the boundaries.

\begin{figure}[tb]
  \centering
  \includegraphics[width=\linewidth, clip]{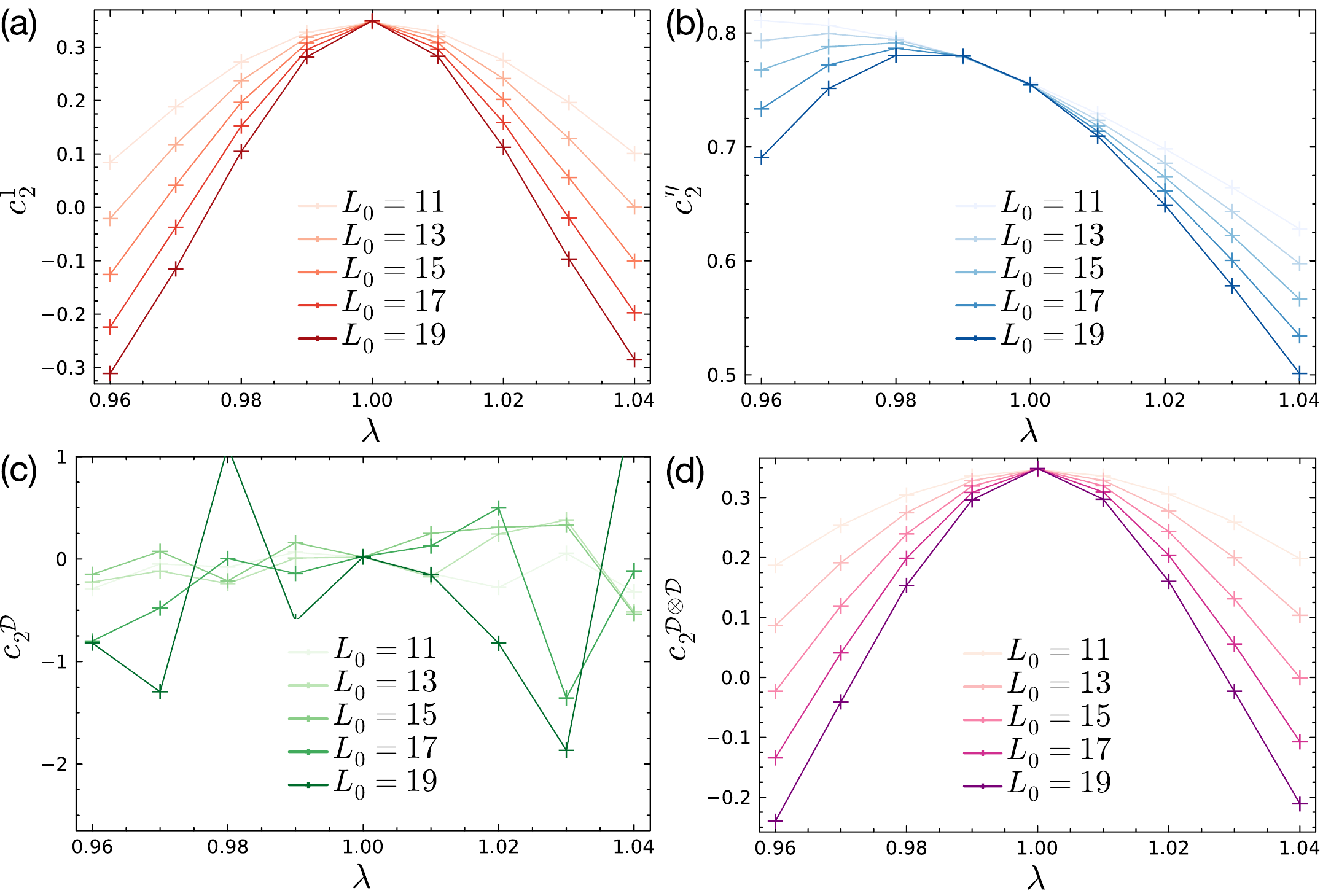}
  \caption{Size-independent term $c_{2}^\mathcal{A}$ in the $\alpha\,{=}\,2$ SRE of the closed chain with (a)~the identity defect $1$, (b)~the $\mathbb{Z}_2$ defect $\eta$, and (c)~the duality defect $\mathcal{D}$. The numerical data were obtained using the replica-Pauli MPS method~\cite{tarabunga2024nonstabilizernessa}. The data are extracted from the SRE fitted to $M_2=m_2L \,{-}\, c_2^{\mathcal{A}} + r/L$ with $L\in\qty{L_0{-}5,L_0{-}3,\ldots,L_0{+}5}$ and $L_0=11,13,\ldots,19$. The estimated universal values at the critical point $\lambda=1$ are $c_2^1=\ln\sqrt{2}$, $c_2^\eta=0.755(1)$, and $c_2^\mathcal{D}=0.020(3)$.
  (d)~Size-independent term for two duality defects is extracted from the SRE of $L$ qubits fitted to $M_2=m_2(L-1) \,{-}\, c_2^{\mathcal{D}{\otimes}\mathcal{D}}$, accounting for the defect $\mathcal{T}^{-}$. The universal value reads $c_2^{\mathcal{D}\otimes\mathcal{D}} \,{=}\, \ln\sqrt{2}$.}
  \label{fig:sre-topological}
\end{figure}

\textit{Topological defects in the Ising CFT.---}
We next demonstrate that the SRE can also serve as a probe of topological defects and their fusion rules.
In the Ising model, the insertion of a topological defect $\eta$ or $\mathcal{D}$ corresponds to a local modification of the Hamiltonian, realized by $Z_{j}Z_{j+1}\mapsto -Z_{j}Z_{j+1}$ or $(Z_{j-1}Z_{j}\,{+}\,X_j)\mapsto Z_{j-1}X_{j}$, respectively.
The presence of such a defect alters the sector of the CFT and its nonstabilizerness.
This change can be detected by a universal size-independent term in the SRE:
\begin{eqnarray}
  c_{\alpha}^\mathcal{A}=\frac{1}{\alpha-1}\ln g_{\alpha}^\mathcal{A}.
\end{eqnarray}
As shown in Figs.~\ref{fig:sre-topological}(a-c), the numerically estimated values of $c_2^\eta$ and $c_2^\mathcal{D}$ are indeed distinct from the defect-free value~\cite{hoshino2025stabilizer}: $c_2^1=\ln\sqrt{2}$, indicating that this universal quantity can distinguish different sectors.
We remark that the failure of data collapse in Fig.~\ref{fig:sre-topological}(c) should be attributed to the fact that the perturbation by $X_j$ in the off-critical region is no longer relevant in the sector containing the duality defect~\cite{chang2019topological,lin2023duality,shao2024whats}.

The use of the SRE becomes even more useful when applied to systems with multiple defects.
One can successively move and fuse defects until a multidefect system effectively reduces to a single-defect problem.
The fusion rules $\eta \otimes \eta \,{=}\, 1$ and $\eta \otimes \mathcal{D} \,{=}\, \mathcal{D}$, for instance, immediately imply the relations $c_{\alpha}^{\eta\otimes\eta} \,{=}\, c_{\alpha}^{1}$ and $c_{\alpha}^{\eta\otimes\mathcal{D}} \,{=}\, c_{\alpha}^\mathcal{D}$, respectively.
A more subtle case arises from the noninvertible fusion rule of two duality defects, which on the lattice reads $\mathcal{D}\otimes\mathcal{D}=\mathcal{T}^{-}(1\oplus\eta)$.
This fusion leads to a direct sum of two defect channels $1$ and $\eta$ while the accompanying defect $\mathcal{T}^{-}$ removes a site from the chain~\cite{seiberg2024noninvertible}.
In the low-temperature limit $\beta\gg L$, the system selects the sector with the lowest energy, which corresponds to the identity channel $1$ in the present case.
Consequently, the SRE of the $L$-qubit ground state with two $\mathcal{D}$ defects should match that of the defect-free ground state of size $L-1$.
This prediction is supported by our numerical results in Fig.~\ref{fig:sre-topological}(d), where we find that the universal term $c_2^{\mathcal{D}\otimes\mathcal{D}}$ matches the defect-free value $c_2^1$.
These results provide a numerical validation of property (II) postulated above.
In fact, one can determine how many topological/factorizing defects are included in a given state by computing the SRE and using properties (I) and (II), and the additivity of the SRE.

\textit{Discussion.---}
Our results can be elevated to an active discovery tool for fusion rules of topological defects.
The first step in determining fusion rules is the identification of the lattice realizations of topological defects.
To this end, we first constrain the search space of candidate defect realizations based on physical requirements: Hermiticity, compatibility with movement with Clifford unitaries, and the transmissive nature of the topological defects.
The resulting candidates are then classified into equivalence classes under Clifford unitaries to further reduce the search space.
Next, we numerically compute the SRE for each candidate class.
The emergence of a universal size-independent term---distinct from the logarithmic scaling characteristic of factorizing defects---serves as the signature of a topological defect.
Finally, assuming that fusions are also implemented by Clifford unitaries, we apply Clifford unitaries to pairs of adjacent topological defects, and by checking whether it maps to a single defect or decouples a site to form a direct sum of defects, the fusion rules can be determined.
Consequently, our framework provides a systematic and practical tool for discovering topological defects and their algebraic structure in unknown critical spin chains.
We refer the reader to the Supplemental Material~\cite{SM} for a more detailed demonstration of this procedure.

Our findings open several avenues for future investigation.
An immediate extension is to analyze the SRE in other CFTs possessing noninvertible symmetries, such as other minimal models or the compactified free boson.
In the latter, investigating the SRE as a function of the compactification radius may offer a new perspective on T-duality.
One could also explore models with higher central charge, for instance, by leveraging the mapping of the XX model to a doubled Ising CFT~\cite{verresen2021gapless,pace2025lattice,su2025$z_2$}.
From a quantum information perspective, our work suggests that the SRE can quantify the magic resource cost of creating and manipulating topological defects, which might have implications for topological quantum computation.
Extending our framework to two-dimensional systems, particularly those with topological order, could reveal how nonstabilizerness characterizes phases with noninvertible higher-form symmetries.
While our specific results rely on the structure of $(1+1)$D CFT, the appearance of universal data in subleading terms of the SRE is expected to be a more general feature, as observed in previous studies of participation entropy in higher dimensions~\cite{luitz2014universal,luitz2014shannonrenyi,misguich2017finitesize}.
Finally, studying the quench dynamics of the SRE in systems with defects may yield novel insights into the interplay between magic propagation and topological constraints.


\begin{acknowledgments}
  \textit{Acknowledgments---}
  We are grateful to Shunsuke Furukawa and Marcin Kalinowski for useful discussions.
  We especially thank Masaki Oshikawa for collaboration on related works~\cite{hoshino2025stabilizer,hoshino2025entanglement,ashida2024systemenvironment}.
  We used the ITensor package~\cite{fishman2022itensor,fishman2022codebase} for tensor-network calculations.
  M.~H. was supported by FoPM, WINGS Program, the University of Tokyo.
  Y. A. acknowledges support from the Japan Society for the Promotion of Science through Grant No. JP19K23424 and from JST FOREST Program (Grant No. JPMJFR222U, Japan) and JST CREST (Grant No. JPMJCR23I2, Japan).
\end{acknowledgments}

\bibliography{HA2025}

\onecolumngrid
\begin{center}
  \textbf{\Large End Matter}
\end{center}
\beginendmatt

\textit{Fusion of the topological defects in the lattice model.---}
Here, we demonstrate that the movement and fusion of the topological defects in the lattice model can be implemented by Clifford unitaries.
The defect Hamiltonian with the invertible $\mathbb{Z}_2$ defect $\eta$ inserted at the bond $(L,1)$ is written as
\begin{equation}
  H_{\eta}^{(L,1)} = -(-Z_LZ_1 + X_1) - \sum_{j=2}^{L}(Z_{j-1}Z_j+X_j).
\end{equation}
The topological nature of this defect allows us to move the defect by a local unitary $U_{\eta}^j=X_j$, such that
\begin{align}
  H_{\eta}^{(1,2)}
   & = U_{\eta}^1 H_{\eta}^{(L,1)}(U_{\eta}^1)^{-1}\notag \\
   & = -(-Z_1Z_2+X_2)-\sum_{j\neq 2}(Z_{j-1}Z_j+X_j).
\end{align}
The local unitary $U_{\eta}^j$ is called the movement operator, and its action is diagrammatically represented as
\begin{equation}
  U_\eta^j = X_j =
  \begin{tikzpicture}[scale=0.8, baseline=0]
    \foreach \x in {-1.0, 0.0, 1.0} {
        \fill (\x, 0.4) circle (2.5pt);
        \fill (\x, -0.4) circle (2.5pt);
      }

    \draw[dashed, thick] (-2.0, 0.4) -- (-1.0, 0.4);
    \draw[dashed, thick] (-2.0, -0.4) -- (-1.0, -0.4);
    \draw[dashed, thick] (1.0, 0.4) -- (2.0, 0.4);
    \draw[dashed, thick] (1.0, -0.4) -- (2.0, -0.4);
    \draw[thick] (-1.0, 0.4) -- (1.0, 0.4);
    \draw[thick] (-1.0, -0.4) -- (1.0, -0.4);

    \draw[blue, very thick] (-0.5, -1.0) -- (-0.5, 0.0);  
    \draw[blue, very thick] (-0.5, 0.0) -- (0.5, 0.0);    
    \draw[blue, very thick] (0.5, 0.0) -- (0.5, 0.8);     

    \node[below, font=\small] at (-1.0, -0.5) {$j{-}1$};
    \node[below, font=\small] at (0.0, -0.5) {$j$};
    \node[below, font=\small] at (1.0, -0.5) {$j+1$};

    \node[blue, above, font=\small] at (0.5, 0.8) {$\eta$};
    \node[blue, below, font=\small] at (-0.5, -1.0) {$\eta$};
  \end{tikzpicture}.
\end{equation}
We note that, since the movement operator is a Clifford unitary, the $\eta$ defect can be moved without consuming energy or magic.

We now consider the fusion of two $\eta$ defects.
After applying the movement operator, the two defects can be brought to adjacent bonds, resulting in the defect Hamiltonian
\begin{equation}
  H_{\eta;\eta}^{(L,1):(1,2)} = -\sum_{i=1,2}(-Z_{j-1}Z_j+X_j)-\sum_{j=3}^{L}(Z_{j-1}Z_j+X_j).
\end{equation}
The fusion of these defects can be implemented by a unitary $\lambda_{\eta\otimes\eta}^j=X_j$, which acts as
\begin{equation}
  \lambda_{\eta\otimes\eta}^1 H_{\eta;\eta}^{(L,1);(1,2)} (\lambda_{\eta\otimes\eta}^1)^{-1} = H,
\end{equation}
where we recall that $H$ represents the Hamiltonian of the periodic chain in Eq.~\eqref{eq:tfim} in the main text.
This is the lattice realization of the fusion rule $\eta\otimes\eta=1$, which can be diagrammatically represented as
\begin{equation}
  \lambda_{\eta\otimes\eta}^j = X_j =
  \begin{tikzpicture}[scale=0.8, baseline=0]
    \foreach \x in {-1.0, 0.0, 1.0} {
        \fill (\x, 0.4) circle (2.5pt);
        \fill (\x, -0.4) circle (2.5pt);
      }

    \draw[dashed, thick] (-2.0, 0.4) -- (-1.0, 0.4);
    \draw[dashed, thick] (-2.0, -0.4) -- (-1.0, -0.4);
    \draw[dashed, thick] (1.0, 0.4) -- (2.0, 0.4);
    \draw[dashed, thick] (1.0, -0.4) -- (2.0, -0.4);
    \draw[thick] (-1.0, 0.4) -- (1.0, 0.4);
    \draw[thick] (-1.0, -0.4) -- (1.0, -0.4);

    \draw[blue, very thick] (-0.5, -1.0) -- (-0.5, 0.0);  
    \draw[blue, very thick] (-0.5, 0.0) -- (0.5, 0.0);    
    \draw[blue, very thick] (0.5, 0.0) -- (0.5, -1.0);     

    \node[below, font=\small] at (-1.2, -0.5) {$j{-}1$};
    \node[below, font=\small] at (0.0, -0.5) {$j$};
    \node[below, font=\small] at (1.2, -0.5) {$j+1$};

    \node[blue, below, font=\small] at (0.5, -1.0) {$\eta$};
    \node[blue, below, font=\small] at (-0.5, -1.0) {$\eta$};
  \end{tikzpicture}.
\end{equation}

The defect Hamiltonian with the noninvertible duality defect $\mathcal{D}$ inserted at the bond $(L,1)$ is written as
\begin{equation}
  H_{\mathcal{D}}^{(L,1)} = -\sum_{j=2}^{L}(Z_{j-1}Z_j+X_j) -Z_LX_1.
\end{equation}
The movement operator for this defect is $U_{\mathcal{D}}^j=\mathsf{CZ}_{j,j+1}\mathsf{H}_j$, where $\mathsf{H}_j$ is the Hadamard gate and $\mathsf{CZ}_{j,k}$ is the CZ gate.
These unitaries act on the Pauli matrices as
\begin{align}
  \mathsf{H}_j:      & \quad X_j\mapsto Z_j,\; Z_j\mapsto X_j, \\
  \mathsf{CZ}_{j,k}: & \quad \left\{
  \begin{array}{cc}
    X_j\mapsto X_jZ_k,   & Z_j\mapsto Z_j \\
    X_{k}\mapsto X_kZ_j, & Z_k\mapsto Z_k
  \end{array}
  \right..
\end{align}
Then, the diagrammatic representation for the movement operator $U_{\mathcal{D}}^j$ is given by
\begin{equation}
  U_{\mathcal{D}}^j = \mathsf{CZ}_{j,j+1}\mathsf{H}_j =
  \begin{tikzpicture}[scale=0.8, baseline=0]
    \foreach \x in {-1.0, 0.0, 1.0} {
        \fill (\x, 0.4) circle (2.5pt);
        \fill (\x, -0.4) circle (2.5pt);
      }

    \draw[dashed, thick] (-2.0, 0.4) -- (-1.0, 0.4);
    \draw[dashed, thick] (-2.0, -0.4) -- (-1.0, -0.4);
    \draw[dashed, thick] (1.0, 0.4) -- (2.0, 0.4);
    \draw[dashed, thick] (1.0, -0.4) -- (2.0, -0.4);
    \draw[thick] (-1.0, 0.4) -- (1.0, 0.4);
    \draw[thick] (-1.0, -0.4) -- (1.0, -0.4);

    \draw[red, very thick] (-0.5, -1.0) -- (-0.5, 0.0);  
    \draw[red, very thick] (-0.5, 0.0) -- (0.5, 0.0);    
    \draw[red, very thick] (0.5, 0.0) -- (0.5, 0.8);     

    \node[below, font=\small] at (-1.2, -0.5) {$j{-}1$};
    \node[below, font=\small] at (0.0, -0.5) {$j$};
    \node[below, font=\small] at (1.2, -0.5) {$j+1$};

    \node[red, above, font=\small] at (0.5, 0.8) {$\mathcal{D}$};
    \node[red, below, font=\small] at (-0.5, -1.0) {$\mathcal{D}$};
  \end{tikzpicture}.
\end{equation}

To identify the fusion of the $\eta$ and $\mathcal{D}$ defect, we consider the defect Hamiltonian
\begin{equation}
  H_{\eta;\mathcal{D}}^{(L,1);(1,2)} = -(-Z_LZ_1+X_1)-Z_1X_2-\sum_{j=3}^{L}(Z_{j-1}Z_j+X_j).
\end{equation}
Using the Clifford unitary
\begin{equation}
  \lambda_{\eta\otimes\mathcal{D}}^j = Z_jX_{j+1} =
  \begin{tikzpicture}[scale=0.8, baseline=0]
    \foreach \x in {-1.0, 0.0, 1.0} {
        \fill (\x, 0.4) circle (2.5pt);
        \fill (\x, -0.4) circle (2.5pt);
      }

    \draw[dashed, thick] (-2.0, 0.4) -- (-1.0, 0.4);
    \draw[dashed, thick] (-2.0, -0.4) -- (-1.0, -0.4);
    \draw[dashed, thick] (1.0, 0.4) -- (2.0, 0.4);
    \draw[dashed, thick] (1.0, -0.4) -- (2.0, -0.4);
    \draw[thick] (-1.0, 0.4) -- (1.0, 0.4);
    \draw[thick] (-1.0, -0.4) -- (1.0, -0.4);

    \draw[blue, very thick] (-0.5, -1.0) -- (-0.5, 0.0);  
    \draw[blue, very thick] (-0.5, 0.0) -- (0.5, 0.0);    
    \draw[red, very thick] (0.5, -1.0) -- (0.5, 0.8);     

    \node[below, font=\small] at (-1.2, -0.5) {$j{-}1$};
    \node[below, font=\small] at (0.0, -0.5) {$j$};
    \node[below, font=\small] at (1.2, -0.5) {$j+1$};

    \node[red, above, font=\small] at (0.5, 0.8) {$\mathcal{D}$};
    \node[blue, below, font=\small] at (-0.5, -1.0) {$\eta$};
    \node[red, below, font=\small] at (0.5, -1.0) {$\mathcal{D}$};
  \end{tikzpicture},
\end{equation}
we obtain
\begin{equation}
  \lambda_{\eta\otimes\mathcal{D}}^1 H_{\eta;\mathcal{D}}^{(L,1);(1,2)} (\lambda_{\eta\otimes\mathcal{D}}^1)^{-1}
  = H_{\mathcal{D}}^{(1,2)}.
\end{equation}
Thus, the fusion $\eta\otimes\mathcal{D}=\mathcal{D}$ is implemented by the fusion operator $\lambda_{\eta\otimes\mathcal{D}}$.
For the fusion in the opposite direction $\mathcal{D}\otimes\eta=\mathcal{D}$, we consider the defect Hamiltonian
\begin{equation}
  H_{\mathcal{D};\eta}^{(L,1);(1,2)} = -Z_LX_1-(-Z_1Z_2+X_2)-\sum_{j=3}^{L}(Z_{j-1}Z_j+X_j),
\end{equation}
and the fusion operator
\begin{equation}
  \lambda_{\mathcal{D}\otimes\eta}^j = U_{\mathcal{D}}^jX_j =
  \begin{tikzpicture}[scale=0.8, baseline=0]
    \foreach \x in {-1.0, 0.0, 1.0} {
        \fill (\x, 0.4) circle (2.5pt);
        \fill (\x, -0.4) circle (2.5pt);
      }

    \draw[dashed, thick] (-2.0, 0.4) -- (-1.0, 0.4);
    \draw[dashed, thick] (-2.0, -0.4) -- (-1.0, -0.4);
    \draw[dashed, thick] (1.0, 0.4) -- (2.0, 0.4);
    \draw[dashed, thick] (1.0, -0.4) -- (2.0, -0.4);
    \draw[thick] (-1.0, 0.4) -- (1.0, 0.4);
    \draw[thick] (-1.0, -0.4) -- (1.0, -0.4);

    \draw[red, very thick] (-0.5, -1.0) -- (-0.5, 0.0);  
    \draw[red, very thick] (-0.5, 0.0) -- (0.5, 0.0);    
    \draw[red, very thick] (0.5, 0.0) -- (0.5, 0.8);     
    \draw[blue, very thick] (0.5, 0.0) -- (0.5, -1.0);

    \node[below, font=\small] at (-1.2, -0.5) {$j{-}1$};
    \node[below, font=\small] at (0.0, -0.5) {$j$};
    \node[below, font=\small] at (1.2, -0.5) {$j+1$};

    \node[red, above, font=\small] at (0.5, 0.8) {$\mathcal{D}$};
    \node[red, below, font=\small] at (-0.5, -1.0) {$\mathcal{D}$};
    \node[blue, below, font=\small] at (0.5, -1.0) {$\eta$};
  \end{tikzpicture},
\end{equation}
which acts as
\begin{equation}
  \lambda_{\mathcal{D}\otimes\eta}^1 H_{\mathcal{D};\eta}^{(L,1);(1,2)}(\lambda_{\mathcal{D}\otimes\eta}^1)^{-1} = H_{\mathcal{D}}^{(1,2)}.
\end{equation}
Thus, the fusion rule $\eta\otimes\mathcal{D}=\mathcal{D}\otimes\eta=\mathcal{D}$ can also be implemented by a Clifford unitary.

We next consider the fusion of two duality defects $\mathcal{D}$.
The defect Hamiltonian with the defects $\mathcal{D}$ on two adjacent bonds is given by
\begin{equation}\label{eq:hamiltonian_dd}
  H_{\mathcal{D};\mathcal{D}}^{(L,1);(1,2)} = -Z_LX_1-Z_1X_2-\sum_{j=3}^{L}(Z_{j-1}Z_j+X_j).
\end{equation}
The fusion operator in this case is given by $\lambda_{\mathcal{D}\otimes\mathcal{D}}^1=(U_{\mathcal{D}}^1)^{-1}=\mathsf{H}_1\mathsf{CZ}_{1,2}$, acting on the defect Hamiltonian as
\begin{align}
    & \lambda_{\mathcal{D}\otimes\mathcal{D}}^1 H_{\mathcal{D};\mathcal{D}}^{(L,1);(1,2)}(\lambda_{\mathcal{D};\mathcal{D}}^1)^{-1} \notag \\
  = & -\sum_{j=3}^{L}(Z_{j-1}Z_j+X_j)-Z_1Z_LZ_2-X_2.
\end{align}
Since the qubit at site $1$ is now completely decoupled from the rest of the chain, we can write the above defect Hamiltonian as
\begin{equation}
  H_{\mathcal{D};\mathcal{D}}^{(L,1);(1,2)} \stackrel{\lambda_{\mathcal{D}\otimes\mathcal{D}}^1}{\longmapsto} H_{\mathcal{T}^{-}}^1\otimes\ketbra*{0}{0}_1 + H_{\mathcal{T}^{-}\eta}^1\otimes\ketbra*{1}{1}_1,
\end{equation}
where
\begin{align}
  H_{\mathcal{T}^{-}}^1 = -(Z_LZ_2+X_2)-\sum_{j=3}^{L}(Z_{j-1}Z_j+X_j), \\
  H_{\mathcal{T}^{-}\eta}^1 = -(-Z_LZ_2+X_2)-\sum_{j=3}^{L}(Z_{j-1}Z_j+X_j),
\end{align}
are the defect Hamiltonians acting on the Hilbert space with one fewer lattice site.
Thus, the noninvertible fusion rule in the lattice model reads $\mathcal{D}\otimes\mathcal{D}=\mathcal{T}^{-}\,{\oplus}\,\mathcal{T}^{-}\eta$.
We diagrammatically denote the action of the fusion operator as
\begin{equation}
  \lambda_{\mathcal{D}\otimes\mathcal{D}}^1 = \mathsf{H}_1\mathsf{CZ}_{1,2} =
  \begin{tikzpicture}[scale=0.8, baseline=0]
    \foreach \x in {-1.0, 0.0, 1.0} {
        \fill (\x, 0.4) circle (2.5pt);
        \fill (\x, -0.4) circle (2.5pt);
      }

    \draw[dashed, thick] (-2.0, 0.4) -- (-1.0, 0.4);
    \draw[dashed, thick] (-2.0, -0.4) -- (-1.0, -0.4);
    \draw[dashed, thick] (1.0, 0.4) -- (2.0, 0.4);
    \draw[dashed, thick] (1.0, -0.4) -- (2.0, -0.4);
    \draw[thick] (-1.0, 0.4) -- (1.0, 0.4);
    \draw[thick] (-1.0, -0.4) -- (1.0, -0.4);

    \draw[red, very thick] (-0.5, -1.0) -- (-0.5, 0.0);  
    \draw[red, very thick] (-0.5, 0.0) -- (0.5, 0.0);    
    \draw[red, very thick] (0.5, 0.0) -- (0.5, -1.0);     
    \draw[blue, very thick] (0.0, 0.0) -- (0.0, 0.8);

    \node[below, font=\small] at (-1.2, -0.5) {$L$};
    \node[below, font=\small] at (0.0, -0.5) {$1$};
    \node[below, font=\small] at (1.2, -0.5) {$2$};

    \node[red, below, font=\small] at (0.5, -1.0) {$\mathcal{D}$};
    \node[red, below, font=\small] at (-0.5, -1.0) {$\mathcal{D}$};
    \node[blue, above, font=\small] at (0.0, 0.8) {$\mathcal{T}^{-}\oplus\mathcal{T}^{-}\eta$};
  \end{tikzpicture}.
\end{equation}
In a finite-size system, the ground-state energy in the sector $H_{\mathcal{T}^{-}}^1$ is lower than that in the sector $H_{\mathcal{T}^{-}\eta}^1$ by an order of $1/L$.
Thus, the ground state of the defect Hamiltonian $H_{\mathcal{D};\mathcal{D}}^{(L,1);(1,2)}$ is given by $\lambda_{\mathcal{D}\otimes\mathcal{D}}^1(\ket*{0}\,{\otimes}\,\ket*{\psi_{\mathcal{T}^{-}}})$, where $\ket*{\psi_{\mathcal{T}^{-}}}$ is the ground state of $H_{\mathcal{T}^{-}}^1$.
Since the SRE is additive and stable against Clifford unitaries, we obtain the relation $M_{\alpha}(\psi_{\mathcal{D};\mathcal{D}})=M_{\alpha}(\psi_{\mathcal{T}^{-}})$, where $\psi_{\mathcal{D};\mathcal{D}}$ is the ground state of the Hamiltonian with two $\mathcal{D}$ defects (see, e.g., Eq.~\eqref{eq:hamiltonian_dd}).

The fusion between topological defects and boundaries can be constructed in a similar manner.
To focus on one side of the boundary, we consider a semi-infinite chain with the following Hamiltonian:
\begin{equation}
  H_{\ket*{A}} = -\sum_{j=1}^{\infty}(Z_jZ_{j+1} + X_j) + a  Z_1,
\end{equation}
where $A = f,\uparrow,\downarrow$ for $a=0,-1,+1$, respectively.
To see how the topological defect fuses with the boundary, we insert the $\eta$ defect at the bond $(1,2)$, rendering the Hamiltonian
\begin{equation}
  H_{A;\eta}^{(1,2)} = -\sum_{j=2}^{\infty}(Z_jZ_{j+1}+X_j) - (-Z_1Z_2 + X_1) + a Z_1.
\end{equation}
The fusion operator in this case is given as
\begin{equation}
  \lambda_{\eta\ast \ket*{A}} = X_1 =
  \begin{tikzpicture}[scale=0.8, baseline=0]
    \foreach \x in {-1.0, 0.0} {
        \fill (\x, 0.4) circle (2.5pt);
        \fill (\x, -0.4) circle (2.5pt);
      }

    \draw[dotted, thick] (-1.8, 0.4) -- (-1.0, 0.4);
    \draw[dotted, thick] (-1.8, -0.4) -- (-1.0, -0.4);
    \draw[dashed, thick] (0, 0.4) -- (1.0, 0.4);
    \draw[dashed, thick] (0, -0.4) -- (1.0, -0.4);
    \draw[thick] (-1.0, 0.4) -- (0, 0.4);
    \draw[thick] (-1.0, -0.4) -- (0, -0.4);

    \draw[magenta, very thick] (-1.5, -1.0) -- (-1.5, 0.0);
    \draw[blue, very thick] (-0.5, -1.0) -- (-0.5, 0.0);
    \draw[blue, very thick] (-0.5, 0.0) -- (-1.5, 0.0);
    \draw[violet, very thick] (-1.5, 0.0) -- (-1.5, 0.8);

    \node[below, font=\small] at (-1.0, -0.5) {$1$};
    \node[below, font=\small] at (0.0, -0.5) {$2$};

    \node[violet, above, font=\small] at (-1.5, 0.8) {$\eta\ast\ket*{A}$};
    \node[magenta, below, font=\small] at (-1.5, -1.0) {$\ket*{A}$};
    \node[blue, below, font=\small] at (-0.5, -1.0) {$\eta$};
  \end{tikzpicture},
\end{equation}
which acts on the Hamiltonian as
\begin{equation}
  \lambda_{\eta\ast\ket*{A}} H_{A;\eta}^{(1,2)} (\lambda_{\eta\ast\ket*{A}})^{-1}
  = -\sum_{j=1}^{\infty}(Z_jZ_{j+1}+X_j) -aZ_1.
\end{equation}
This operation implements the fusion rule Eq.~\eqref{eq:fusion-boundary-eta} in the main text.

We next consider the fusion between the duality defect $\mathcal{D}$ and the boundary.
First, for the spin-up boundary $\ket*{\uparrow}$, the Hamiltonian
\begin{equation}
  H_{\ket*{\uparrow};\mathcal{D}}^{(1,2)} = -\sum_{j=3}^{\infty}(Z_{j-1}Z_j+X_j)-Z_1X_2-X_1 -Z_1
\end{equation}
transforms as
\begin{equation}
  \lambda_{\mathcal{D}\ast\ket*{\uparrow}} H_{\ket*{\uparrow};\mathcal{D}}^{(1,2)} (\lambda_{\mathcal{D}\ast\ket*{\uparrow}})^{-1} = H_{\ket*{f}}
\end{equation}
under the fusion operator
\begin{equation}
  \lambda_{\mathcal{D}\ast \ket*{\uparrow}} = (U_{\mathcal{D}}^1)^{-1} = \mathsf{H}_1\mathsf{CZ}_{1,2} =
  \begin{tikzpicture}[scale=0.8, baseline=0]
    \foreach \x in {-1.0, 0.0} {
        \fill (\x, 0.4) circle (2.5pt);
        \fill (\x, -0.4) circle (2.5pt);
      }

    \draw[dotted, thick] (-1.8, 0.4) -- (-1.0, 0.4);
    \draw[dotted, thick] (-1.8, -0.4) -- (-1.0, -0.4);
    \draw[dashed, thick] (0, 0.4) -- (1.0, 0.4);
    \draw[dashed, thick] (0, -0.4) -- (1.0, -0.4);
    \draw[thick] (-1.0, 0.4) -- (0, 0.4);
    \draw[thick] (-1.0, -0.4) -- (0, -0.4);

    \draw[magenta, very thick] (-1.5, -1.0) -- (-1.5, 0.0);
    \draw[red, very thick] (-0.5, -1.0) -- (-0.5, 0.0);
    \draw[red, very thick] (-0.5, 0.0) -- (-1.5, 0.0);
    \draw[violet, very thick] (-1.5, 0.0) -- (-1.5, 0.8);

    \node[below, font=\small] at (-1.0, -0.5) {$1$};
    \node[below, font=\small] at (0.0, -0.5) {$2$};

    \node[violet, above, font=\small] at (-1.5, 0.8) {$\ket*{f}$};
    \node[magenta, below, font=\small] at (-1.5, -1.0) {$\ket*{\uparrow}$};
    \node[red, below, font=\small] at (-0.5, -1.0) {$\mathcal{D}$};
  \end{tikzpicture},
\end{equation}
which implements $\mathcal{D}\,{\ast}\ket*{\uparrow}=\ket*{f}$.
Similarly, the fusion operator for the spin-down boundary $\ket*{\downarrow}$ is given by $\lambda_{\mathcal{D}\ast\ket*{\downarrow}}=\lambda_{\mathcal{D}\ast\ket*{\uparrow}}X_1$.
Finally, the fusion between the duality defect $\mathcal{D}$ and the free boundary $\ket*{f}$ can be described by the Hamiltonian
\begin{equation}
  H_{\ket*{f};\mathcal{D}}^{(1,2)} = -\sum_{j=3}^{\infty}(Z_{j-1}Z_j+X_j)-Z_1X_2-X_1.
\end{equation}
Using the fusion operator $\lambda_{\mathcal{D}\ast\ket*{f}}=(U_{\mathcal{D}}^1)^{-1}$, we obtain
\begin{align}
    & \lambda_{\mathcal{D}\ast\ket*{f}} H_{\ket*{f};\mathcal{D}}^{(1,2)} (\lambda_{\mathcal{D}\ast\ket*{f}})^{-1}\notag        \\
  = & -Z_1Z_2-\sum_{j=2}^{\infty}(Z_jZ_{j+1}+X_j)\notag                                                                        \\
  = & H_{\mathcal{T}^{-}\ket*{\uparrow}}\otimes\ketbra*{0}{0}_1 + H_{\mathcal{T}^{-}\ket*{\downarrow}}\otimes\ketbra*{1}{1}_1,
\end{align}
where
\begin{align}
  H_{\mathcal{T}^{-}\ket*{\uparrow}} = -Z_2-\sum_{j=2}^{\infty}(Z_jZ_{j+1}+X_j), \\
  H_{\mathcal{T}^{-}\ket*{\downarrow}} = +Z_2-\sum_{j=2}^{\infty}(Z_jZ_{j+1}+X_j),
\end{align}
are the Hamiltonians with spin-up and -down boundaries with one fewer lattice site.
This explains how the superposition $\ket*{\uparrow}+\ket*{\downarrow}$ of the boundary is realized in the lattice model.
The diagrammatic representation for this fusion is given by
\begin{equation}
  \lambda_{\mathcal{D}\ast \ket*{f}} = (U_{\mathcal{D}}^1)^{-1} = \mathsf{H}_1\mathsf{CZ}_{1,2} =
  \begin{tikzpicture}[scale=0.8, baseline=0]
    \foreach \x in {-1.0, 0.0} {
        \fill (\x, 0.4) circle (2.5pt);
        \fill (\x, -0.4) circle (2.5pt);
      }

    \draw[dotted, thick] (-1.8, 0.4) -- (-1.0, 0.4);
    \draw[dotted, thick] (-1.8, -0.4) -- (-1.0, -0.4);
    \draw[dashed, thick] (0, 0.4) -- (1.0, 0.4);
    \draw[dashed, thick] (0, -0.4) -- (1.0, -0.4);
    \draw[thick] (-1.0, 0.4) -- (0, 0.4);
    \draw[thick] (-1.0, -0.4) -- (0, -0.4);

    \draw[magenta, very thick] (-1.0, 0.0) -- (-1.0, 0.8);
    \draw[red, very thick] (-0.5, -1.0) -- (-0.5, 0.0);
    \draw[red, very thick] (-0.5, 0.0) -- (-1.0, 0.0);
    \draw[violet, very thick] (-1.5, 0.0) -- (-1.0, 0.0);
    \draw[violet, very thick] (-1.5, -1.0) -- (-1.5, 0.0);

    \node[below, font=\small] at (-1.0, -0.5) {$1$};
    \node[below, font=\small] at (0.0, -0.5) {$2$};

    \node[magenta, above, font=\small] at (-1.0, 0.8) {$\ket*{\uparrow}{+}\ket*{\downarrow}$};
    \node[violet, below, font=\small] at (-1.5, -1.0) {$\ket*{f}$};
    \node[red, below, font=\small] at (-0.5, -1.0) {$\mathcal{D}$};
  \end{tikzpicture}.
\end{equation}
Altogether, we have demonstrated that all the fusions of the boundaries and topological defects in the Ising CFT can be implemented by Clifford unitaries.

\onecolumngrid
\clearpage
\begin{center}
  \textbf{\large Supplemental Material: Stabilizer R\'{e}nyi Entropy Encodes Fusion Rules of Topological Defects and Boundaries}
\end{center}
\beginsupplement

We provide the details of our BCFT analysis in obtaining the universal properties of the stabilizer R\'{e}nyi entropy (SRE) in the presence of conformal defects.
Specifically, we explicitly compute the scaling dimensions of the boundary condition changing operators (BCCOs) necessary to determine the universal terms contained in the SRE of the critical Ising model with open boundaries.
We first construct conformal boundary states of $(N=2\alpha)$-component $S^1/\mathbb{Z}_2$ free-boson CFT which are consistent with the following boundary conditions $\Gamma_{1,2}$~\cite{hoshino2025stabilizer} (see also Fig.~\ref{fig:partition-function}(a) in the main text):
\begin{equation}\label{eq:gamma1_description}
  \Gamma_1:\left\{
  \begin{array}{lc}
    \displaystyle\sum_{i=1}^{2\alpha} \phi_i & \text{NBC} \\
    \phi_1-\phi_2=0                          & \text{DBC} \\
    \phi_2-\phi_3=0                          & \text{DBC} \\
    \quad\vdots                              &            \\
    \phi_{2\alpha-1} - \phi_{2\alpha}=0      & \text{DBC}
  \end{array}
  \right.,
\end{equation}
and
\begin{equation}\label{eq:gamma2_description}
  \Gamma_2:\left\{
  \begin{array}{lc}
    \phi_i-\phi_{i+\alpha}=0 \quad(i=1,2,\ldots,\alpha) & \text{DBC} \\
    \phi_i+\phi_{i+\alpha} \quad(i=1,2,\ldots,\alpha)   & \text{NBC} \\
  \end{array}
  \right.,
\end{equation}
where NBC and DBC stand for Neumann and Dirichlet boundary conditions, respectively.
Then, using the known correspondance between the boundary states in the double Ising CFT and the $S^1/\mathbb{Z}_2$ free-boson CFT~\cite{oshikawa1997boundary}, we compute the amplitudes between $\Gamma_{1,2}$ and $\Gamma_{A/B}$ to extract the scaling dimension of the BCCOs.

\subsection{Construction of the boundary states in multicomponent \texorpdfstring{$S^1/\mathbb{Z}_2$}{orbifold} free-boson CFT}
To make the material self-contained, we here review the construction of a class of boundary states in the multicomponent $S^1/\mathbb{Z}_2$ free-boson CFT.
Our construction builds upon the boundary states in the multicomponent $S^1$ free boson, from which we derive the corresponding boundary states in the multicomponent $S^1/\mathbb{Z}_2$ free boson through symmetrization.

The Lagrangian density of the bulk theory is
\begin{equation}
  \mathcal{L} = \frac{1}{2\pi}(\partial_\mu\vec{\phi})^2\quad (\mu=\tau,x),
\end{equation}
where $\vec{\phi}=(\phi_1,\phi_2,\ldots,\phi_N)$ is the $N$-component boson field.
We describe the dual field of $\vec{\phi}$ as $\vec{\theta}=(\theta_1,\theta_2,\ldots,\theta_N)$.
These boson fields are compactified as
\begin{equation}
  \vec{\phi}\sim\vec{\phi}+2\pi\vec{R},\quad \vec{\theta}\sim\vec{\theta}+2\pi\vec{K}.
\end{equation}
The vectors $\vec{R}$ and $\vec{K}$ belong to the following compactification lattice,
\begin{equation}
  \Lambda = \qty{\vec{x}\;|\; \vec{x}=\sum_{i=1}^{N}n_i\vec{e}_i,\;n_i\in\mathbb{Z}},\quad
  \Lambda^\ast = \qty{\vec{y}\;|\; \vec{y}=\sum_{i=1}^{N}m_i\vec{f}_i,\;m_i\in\mathbb{Z}},\quad
  \vec{e}_i\cdot\vec{f}_j = \delta_{ij},
\end{equation}
respectively, where the lattice $\Lambda^\ast$ is the dual of $\Lambda$, and $\vec{R}\cdot\vec{K}\in\mathbb{Z}$ is satisfied.
For instance, when all the field components are decoupled, the compactification lattice is simply a square lattice.
The fields $\vec{\phi}$ and $\vec{\theta}$ defined on the cylinder of circumference $L$ can be expanded in the Fourier modes as follows:
\begin{align}
  \vec{\phi}(x,t)
   & = \vec{\phi_0} + \frac{2\pi}{L}\qty(\vec{R}x + \frac{\vec{K}}{2}t)                                              + \frac{1}{2}\sum_{n=1}^\infty\frac{1}{\sqrt{n}}\qty(\vec{a}_n^L e^{-ik_n(x+t)} + \vec{a}_n^R e^{ik_n(x-t)} + \text{H.c.}), \\
  \vec{\theta}(x,t)
   & = \vec{\theta_0} + \frac{2\pi}{L}\qty(\vec{K}x + 2\vec{R}t)                                                           + \sum_{n=1}^\infty\frac{1}{\sqrt{n}}\qty(\vec{a}_n^L e^{-ik_n(x+t)} - \vec{a}_n^R e^{ik_n(x-t)} + \text{H.c.}),
\end{align}
where $\vec{\phi}_0$ and $\vec{\theta}_0$ are the zero-mode operators, $\vec{a}_n^{L(R)}$ is a vector of the annihilation operators of left- (right-) moving oscillator modes having quantum number $n$, and $k_n=2\pi n/L$.
These operators satisfy the following commutation relations:
\begin{equation}
  [(\vec{a}_n^s)_i,(\vec{a}_m^t)_j^\dag]  = \delta_{nm}\delta_{st}\delta_{ij}\;\;(s,t\in\qty{L,R}), \quad
  [(\vec{\phi}_0)_i,(\vec{K})_j]          = i\delta_{ij},                                           \quad
  [(\vec{\theta}_0)_i,(\vec{R})_j]        = i\delta_{ij}.
\end{equation}
The Hamiltonian on the cylinder can be expressed in terms of these operators as follows:
\begin{equation}
  H = \frac{2\pi}{L}\qty(\vec{R}^2 {+} \frac{\vec{K}^2}{4} {+} \sum_{n=1}^\infty n\qty[(\vec{a}_n^L)^\dag{\cdot} \vec{a}_n^L {+} (\vec{a}_n^R)^\dag{\cdot}\vec{a}_n^R] {-} \frac{N}{12}).
\end{equation}
Its ground state is the simultaneous eigenstate of the winding modes $\vec{R},\vec{K}$ and the oscillators $\vec{a}_n^{L/R}$ with zero eigenvalues, and the ground-state energy is the Casimir energy $E_{\mathrm{GS}}(L) = -\pi N/(6L)$.

The condition of conformal invariance of a boundary state $\ket*{\Gamma}$ is given by
\begin{equation}\label{eq:conformal-invariance}
  (L_m - \bar{L}_{-m})\ket*{\Gamma}=0,
\end{equation}
where $L_m,\bar{L}_m$ are the Virasoro generators, expressed in terms of the Fourier modes as follows:
\begin{equation}\label{eq:virasoro-generator}
  L_m = \frac{1}{2}\sum_l :\vec{\alpha}_{m-l}^L\cdot \vec{\alpha}_l^L:,\quad\bar{L}_m = \frac{1}{2}\sum_l :\vec{\alpha}_{m-l}^R\cdot\vec{\alpha}_l^R:.
\end{equation}
Here, $:{\cdots}:$ is the operator normal ordering, and the operators $\vec{\alpha}_n^{L,R}$ are defined as
\begin{equation}
  \vec{\alpha}_n^L  = \left\{\begin{array}{cc}
    -i\sqrt{n}\vec{a}_n^L                     & (n>0) \\\\
    \displaystyle\frac{1}{2}\vec{K} + \vec{R} & (n=0) \\\\
    i\sqrt{n}(\vec{a}_{-n}^L)^\dag            & (n<0)
  \end{array}\right.,
  \quad\quad
  \vec{\alpha}_n^R  = \left\{\begin{array}{cc}
    -i\sqrt{n}\vec{a}_n^R                     & (n>0) \\\\
    \displaystyle\frac{1}{2}\vec{K} - \vec{R} & (n=0) \\\\
    i\sqrt{n}(\vec{a}_{-n}^R)^\dag            & (n<0)
  \end{array}\right..
\end{equation}
A sufficient condition to satisfy Eq.~\eqref{eq:conformal-invariance} is
\begin{equation}\label{eq:sufficient-condition}
  (\vec{\alpha}_m^L - \mathcal{R}\vec{\alpha}_{-m}^R)\ket{\Gamma} = 0,
\end{equation}
where $\mathcal{R}$ is an $N\times N$ orthogonal matrix.
This matrix $\mathcal{R}$ becomes symmetric if each of the components satisfies either the DBC or the NBC.
The states satisfying the sufficient condition~\eqref{eq:sufficient-condition} for $m\neq0$ can be given by the coherent states
\begin{equation}\label{coherent}
  S(\mathcal{R})\ket*{\vec{R},\vec{K}},
\end{equation}
where
\begin{equation}
  S(\mathcal{R}) = \exp[-\sum_{n=1}^{\infty}(\vec{a}_n^L)^\dag\cdot\mathcal{R}(\vec{a}_n^R)^\dag]
\end{equation}
is the squeezing operator, and $\ket*{\vec{R},\vec{K}}$ are the oscillator vacua being the eigenstates of the winding modes $\vec{R},\vec{K}$.
These coherent states are known as the Ishibashi states in BCFT~\cite{ishibashi1989boundary,recknagel2013boundary}.
The condition~\eqref{eq:sufficient-condition} for $m=0$ restricts the allowed states $\ket*{\vec{R},\vec{K}}$ to satisfy
\begin{equation}\label{eq:winding-condition}
  \vec{K} + 2 \vec{R} = \mathcal{R}(\vec{K} - 2\vec{R}).
\end{equation}
Thus, any linear combination of the coherent states $S(\mathcal{R})\ket*{\vec{R},\vec{K}}$ that satisfy~\eqref{eq:winding-condition} are conformally invariant.

To correctly describe the theory with boundaries, physical boundary states must also satisfy the consistency condition known as the Cardy's consistency condition.
To explain this, consider the transition amplitude between a pair of boundary states $\ket*{\mathcal{A}},\ket*{\mathcal{B}}$ as follows:
\begin{equation}
  Z_{\mathcal{A}\mathcal{B}}(q)=\mel*{\mathcal{B}}{e^{-\frac{\beta}{2}H}}{\mathcal{A}},
\end{equation}
where $q=e^{2\pi i\tau}$ with $\tau=i\beta/L$ being the modular parameter.
Under the modular transformation $\tau\to-1/\tau\;(q\to\tilde{q}=e^{-2\pi i/\tau})$, which exchanges the direction of imaginary time and space, this amplitude is written as the trace over the Hilbert space $\mathcal{H}_{\mathcal{A}\mathcal{B}}$ of the theory on a strip with boundary conditions $\mathcal{A}$ and $\mathcal{B}$:
\begin{equation}
  Z_{\mathcal{A}\mathcal{B}}(q) = \Tr e^{-LH_{\mathcal{A}\mathcal{B}}},
\end{equation}
where $H_{\mathcal{A}\mathcal{B}}$ is the corresponding CFT Hamiltonian.
Given that the boundary conditions are conformally invariant, this Hilbert space decomposes into irreducible representations $\mathcal{H}_h$ of the Virasoro algebra.
Thus, the amplitude must be written as
\begin{equation}
  Z_{\mathcal{A}\mathcal{B}}(q)=\sum_{h}n_{\mathcal{A}\mathcal{B}}^{h}\chi_h(\tilde{q}),
\end{equation}
where
\begin{equation}
  n_{\mathcal{A}\mathcal{B}}^{h}\in\mathbb{N}_0\label{cardy}
\end{equation}
is a nonnegative integer interpreted as the number of primary fields with conformal weight $h$ in the spectrum, and $\chi_h(\tilde{q})$ is the Virasoro character.
When $\mathcal{A}=\mathcal{B}$ ($\mathcal{A}\neq\mathcal{B}$), we call the condition~\eqref{cardy} self (mutual) consistency condition.
If the ground state of $H_{\mathcal{A}\mathcal{A}}$ is unique, the boundary condition $\mathcal{A}$ satisfies $n_{\mathcal{A}\mathcal{A}}^0=1$.
We note that the consistency conditions cannot be satisfied by a single coherent state in Eq.~\eqref{coherent} alone.
Thus, to construct physical boundary states, we must take appropriate linear combinations of the coherent states.

We now construct a class of the consistent boundary states for general mixed Dirichlet-Neumann boundary conditions.
To begin with, we consider the simplest case of the fully DBC or fully NBC boundary states.
The fully DBC corresponds to taking $\mathcal{R}=I$, and Eq.~\eqref{eq:winding-condition} translates to $\vec{R}=\vec{0}$.
The consistent boundary state is then
\begin{equation}\label{fullyD}
  \ket*{D(\vec{\phi}_D)}_c := g_D \sum_{\vec{K}\in\Lambda^\ast} e^{-i\vec{K}\cdot\vec{\phi}_D}S(I)\ket*{\vec{0},\vec{K}},
\end{equation}
where the coefficient $e^{-i\vec{K}\cdot\vec{\phi}_D}$ is introduced to ensure that this state is an eigenstate of $\vec{\phi}$ with eigenvalue $\vec{\phi}_D$.
The subscript in $\ket*{\cdots}_c$ stands for \textit{circle} of the $S^1$ theory, which we use to distinguish it from the boundary states in the $S^1/\mathbb{Z}_2$ orbifold theory.
The overall coefficient $g_D$ plays the role of the $g$-factor, which can be given by the overlap of the boundary state with the ground state $\ket*{\mathrm{GS}}=\ket*{\vec{0},\vec{0}}$.

Similarly, the fully NBC corresponds to taking $\mathcal{R}={-}I$, for which the boundary state is
\begin{equation}
  \ket*{N(\vec{\theta}_D)}_c := g_N \sum_{\vec{R}\in\Lambda} e^{-i\vec{R}\cdot\vec{\theta}_D} S(-I)\ket*{\vec{R},\vec{0}},
\end{equation}
where $g_N$ is the corresponding $g$-factor.
Since the NBC for $\vec{\phi}$ is equivalent to DBC for its dual $\vec{\theta}$, these states are labeled by the eigenvalue $\vec{\theta}_D$ of $\vec{\theta}$.
The $g$-factors $g_D$ and $g_N$ are determined from the self-consistency condition and given by
\begin{equation}\label{eq:$g$-factor-dbc}
  g_D = 4^{-N/4}v_0(\Lambda)^{-1/2},            \quad
  g_N = v_0(\Lambda)^{1/2},
\end{equation}
where $v_0(\Lambda)$ is the unit-cell volume of the lattice $\Lambda$.

A general boundary state for mixed Dirichlet-Neumann boundary conditions can be expressed as
\begin{equation}\label{eq:boundary-state-general}
  \ket*{\mathcal{R}(\vec{\phi}_D,\vec{\theta}_D)}_c
  :=g_\mathcal{R} \sum_{\vec{R}\in\Lambda_\mathcal{R}}\sum_{\vec{K}\in\Lambda^\ast_\mathcal{R}} e^{-i\vec{R}\cdot\vec{\theta}_D-i\vec{K}\cdot\vec{\phi}_D}S(\mathcal{R})\ket*{\vec{R},\vec{K}}.
\end{equation}
Here, $\Lambda_\mathcal{R}$ and $\Lambda_\mathcal{R}^\ast$ are subspaces of $\Lambda$ and $\Lambda^\ast$ that satisfy Eq.~\eqref{eq:winding-condition}.
When each of the components satisfies either the DBC or the NBC, we can write $\mathcal{R}=\mathcal{P}_D-\mathcal{P}_N$, where $\mathcal{P}_{D/N}$ is the projection matrix onto the subspace $\mathcal{V}_{D/N}$ satisfying DBC/NBC.
In terms of the projection matrices, Eq.~\eqref{eq:winding-condition} becomes
\begin{equation}
  \mathcal{P}_N\vec{K}=0,\;\mathcal{P}_D\vec{R}=0,
\end{equation}
and the subspaces $\Lambda_\mathcal{R}$ and $\Lambda_\mathcal{R}^\ast$ can be expressed as
\begin{equation}
  \Lambda_\mathcal{R} = \Lambda\cap\mathcal{V}_N,\;\Lambda_\mathcal{R}^\ast = \Lambda^\ast\cap\mathcal{V}_D.
\end{equation}
We note that the $g$-factor $g_{\mathcal{R}}$ does not depend on the zero-mode phases $\vec{\phi}_D,\vec{\theta}_D$.

We now focus on the mixed boundary condition $\Gamma_1$ in Eq.~\eqref{eq:gamma1_description} and construct the corresponding boundary state $\ket*{\Gamma_1}_c$ of the $S^1$ theory.
In this case, the subspace $\mathcal{V}_N$ is a one-dimensional space spanned by the vector $\vec{d}=(1,1,\ldots,1)^{\mathrm{T}}$, which corresponds to the NBC of the center of mass field $\sum_{i=1}^{N}\phi_i$.
The complement of $\mathcal{V}_N$ is the subspace $\mathcal{V}_D$, which corresponds to the DBC of the phase differences $\phi_i-\phi_{i{+}1}$.
The zero-mode phases are simply given by $\vec{\phi}_D=0$ and $\vec{\theta}_D=0$.
Since all the field components are decoupled in the bulk theory, the compactification lattice $\Lambda$ is a square lattice of lattice constant $R$, and the sublattices are written as
\begin{equation}
  \Lambda_\mathcal{R}       = \qty{nR\vec{d}\;\vert\;n\in\mathbb{Z}},  \quad \Lambda_\mathcal{R}^\ast  = \qty{\vec{K}\in\Lambda^\ast \;\vert\;\vec{d}\cdot\vec{K}=0}.
\end{equation}
To calculate the $g$-factor of $\ket*{\Gamma_1}_{c}$, we consider the mutual consistency with the boundary state $\ket*{D(\vec{\phi}_D)}_c$ in Eq.~\eqref{fullyD}.
The amplitude between the two states reads
\begin{equation}\label{eq:d_gamma1_amplitude}
  Z_{D\Gamma_1}(q)
  = {}_c\!\mel*{D(\vec{\phi}_D)}{e^{-\frac{\beta}{2}H}}{\Gamma_1}_{c}
  = \frac{\sqrt{2}\,g_Dg_1^{\mathrm{circ}}}{(\eta(q))^{N-1}}\sqrt{\frac{\eta(q)}{\theta_2(q)}}\sum_{\vec{K}\in\Lambda_\mathcal{R}^\ast} e^{i\vec{K}\cdot\vec{\phi}_D}q^{\vec{K}^2/8},
\end{equation}
where we express the $g$-factor of $\ket*{\Gamma_1}_{c}$ by $g_1^{\mathrm{circ}}$ and use the Dedekind eta function $\eta(q)$ and the theta function $\theta_2(q)$.
We define the eta function and the theta functions as follows:
\begin{align}
  \eta(q)
   & = q^{1/24}\prod_{n=1}^{\infty}(1 - q^n),                                                \\
  \theta_2(q)
   & = \sum_{n\in\mathbb{Z}} q^{(n+1/2)^2/2} = 2q^{1/8}\prod_{n=1}^{\infty}(1-q^n)(1+q^n)^2, \\
  \theta_3(q)
   & = \sum_{n\in\mathbb{Z}}q^{n^2/2} = \prod_{n=1}^{\infty}(1-q^n)(1+q^{n-1/2})^2,          \\
  \theta_4(q)
   & = \sum_{n\in\mathbb{Z}}(-1)^n q^{n^2/2} = \prod_{n=1}^{\infty}(1-q^n)(1-q^{n-1/2})^2.
\end{align}
We note that the contribution $z_{DN}:=\sqrt{\eta(q)/\theta_2(q)}$ originates from the amplitude of the center of mass field, and it is equivalent to the amplitude between the Dirichlet and Neumann boundary states of the single-component $S^1$ theory.
By modular transformation $\tau\to-1/\tau$, the amplitude~\eqref{eq:d_gamma1_amplitude} is rewritten in terms of $\tilde{q}$ as
\begin{equation}\label{eq:d_gamma1_amplitude_tilde}
  Z_{D\Gamma_1}(q)
  = \frac{\sqrt{2}\,g_Dg_1^{\mathrm{circ}}\times 4^{(N-1)/2}}{v_0(\Lambda^\ast_\mathcal{R})(\eta(\tilde{q}))^{N-1}}\sqrt{\frac{\eta(\tilde{q})}{\theta_4(\tilde{q})}}\sum_{\vec{R}\in\tilde{\Lambda}_\mathcal{R}} \tilde{q}^{2\qty(\vec{R}-\frac{\vec{\phi}_D}{2\pi})^2},
\end{equation}
where $\tilde{\Lambda}_{\mathcal{R}}$ is the $(N-1)$-dimensional lattice dual to $\Lambda_{\mathcal{R}}^\ast$.
The mutual consistency leads to the following condition on the $g$-factors:
\begin{equation}
  g_Dg_1^{\mathrm{circ}}\frac{\sqrt{2}\times 4^{(N-1)/2}}{v_0(\Lambda^\ast_\mathcal{R})}=1.
\end{equation}
Thus, the amplitude~\eqref{eq:d_gamma1_amplitude_tilde} can be written as the sum of Virasoro characters as follows:
\begin{equation}
  Z_{D\Gamma_1}(q) = \frac{1}{(\eta(\tilde{q}))^N}\qty(\sum_{n=1}^{\infty}\tilde{q}^{\frac{1}{4}\qty(n-\frac{1}{2})^2})\qty(\sum_{\vec{R}\in\tilde{\Lambda}_{\mathcal{R}}} \tilde{q}^{2\qty(\vec{R} - \frac{\vec{\phi}_D}{2\pi})^2}).
\end{equation}
Here, we used the identity
\begin{equation}
  \sqrt{\frac{\eta(\tilde{q})}{\theta_4(\tilde{q})}} = \frac{\theta_2(\tilde{q}^{1/2})}{2\eta(\tilde{q})} = \frac{1}{\eta(\tilde{q})}\sum_{n=1}^{\infty}\tilde{q}^{\frac{1}{4}\qty(n-\frac{1}{2})^2}.
\end{equation}

We are now in a position to construct the boundary states in the multicomponent $S^1/\mathbb{Z}_2$ free-boson CFT.
To this end, we symmetrize the boundary states of the $S^1$ theory constructed above so that the resulting states are invariant under the $\mathbb{Z}_2$ transformation: $\phi\to-\phi$.
In the case of an $N$-component theory, the $\mathbb{Z}_2$ transformations form a group $G$ with $\abs{G}=2^N$ elements, whose matrix representation in the $N$-dimensional vector space reads $a=\mathrm{diag}(\pm1,\pm1\ldots,\pm1)$ for $a\in G$.
The action of the transformation $a$ on the boundary state is expressed as a unitary transformation $D(a)$ in the following manner
\begin{align}
  D(a)\ket*{\mathcal{R}(\vec{\phi}_D,\vec{\theta}_D)}_c
   & = g_\mathcal{R}\sum_{\vec{R}\in\Lambda_\mathcal{R}}\sum_{\vec{K}\in\Lambda_\mathcal{R}^\ast} e^{-i\vec{R}\cdot\vec{\theta}_D-i\vec{K}\cdot\vec{\phi}_D} S(a\mathcal{R}a)\ket*{a\vec{R},a\vec{K}}\notag      \\
   & = g_\mathcal{R}\sum_{\vec{R}\in a\Lambda_\mathcal{R}}\sum_{\vec{K}\in a\Lambda_\mathcal{R}^\ast} e^{-i\vec{R}\cdot a\vec{\theta}_D-i\vec{K}\cdot a\vec{\phi}_D}S(a\mathcal{R}a)\ket*{\vec{R},\vec{K}}\notag \\
   & = \ket*{a\mathcal{R}a(a\vec{\phi}_D,a\vec{\theta}_D)}_c,
\end{align}
where we use $g_\mathcal{R}=g_{a\mathcal{R}a}$ and $a\Lambda_\mathcal{R}=\Lambda_{a\mathcal{R}a}$ to obtain the last line.

We start from the simplest case of the single-component ($N=1$) $S^1/\mathbb{Z}_2$ free-boson CFT~\cite{oshikawa1997boundary}.
In this case, the orthogonal matrix $\mathcal{R}=\pm 1$ remains invariant under any $a\in G=\qty{\pm 1}$, and the zero modes transform as $\phi_D\to-\phi_D,\theta_D\to-\theta_D$ under the action of $a=-1$.
Thus, when the zero-mode parameters $\phi_D$ and $\theta_D$ take generic values, the symmetrized boundary states for $\mathcal{R}=\pm 1$ in the single-component $S^1/\mathbb{Z}_2$ theory are given by
\begin{equation}\label{sDirichlet}
  \ket*{D(\phi_D)}_{\mathrm{orb}}   = \frac{1}{\sqrt{2}}\qty(\ket*{D(\phi_D)}_c + \ket*{D(-\phi_D)}_c),                  \quad
  \ket*{N(\theta_D)}_{\mathrm{orb}} = \frac{1}{\sqrt{2}}\qty(\ket*{N(\theta_D)}_c + \ket*{N(-\theta_D)}_c).
\end{equation}
The coefficient $1/\sqrt{2}$ is determined from the self-consistency condition.

Meanwhile, when the zero-mode parameters take the fixed-point values of the transformation $G$, i.e., $\phi_D=\phi_E\in\qty{0,\pi R}$ or $\theta_D=\theta_E\in\qty{0,\pi/R}$, the boundary states $\ket*{D(\phi_E)}_c,\ket*{D(-\phi_E)}_c$ and $\ket*{N(\theta_E)}_c,\ket*{N(-\theta_E)}_c$ coincide  and cannot satisfy the consistency condition.
To address this issue, one can introduce the boundary states in the twisted sector, i.e., the $S^1$ free-boson CFT with anti-periodic boundary condition $\phi(x+L,t)=-\phi(x,t)$ on the cylinder.
The boundary states in the twisted sector are labeled by the fixed-point values $\phi_E,\theta_E$, which are the only allowed eigenvalues due to the anti-periodicity.
We denote the boundary states in the twisted sector by the subscript in $\ket*{\cdots}_t$.
Accordingly, the consistent boundary states within the $S^1/\mathbb{Z}_2$ theory can be constructed by the following combination of the fixed-point boundary states in the untwisted and twisted sectors:
\begin{equation}
  \ket*{D(\phi_E)}_{\mathrm{orb}}   = \frac{1}{\sqrt{2}}\ket*{D(\phi_E)}_c \pm 2^{-1/4}\ket*{D(\phi_E)}_t,   \quad
  \ket*{N(\theta_E)}_{\mathrm{orb}} = \frac{1}{\sqrt{2}}\ket*{N(\theta_E)}_c \pm 2^{-1/4}\ket*{N(\theta_E)}_t.
\end{equation}
Here, the coefficient $1/\sqrt{2}$ for the untwisted sector is determined from the mutual consistency with the boundary states in Eq.~\eqref{sDirichlet}, and the coefficient $2^{-1/4}$ for the twisted sector is determined from the self-consistency.

We next consider a multicomponent theory with $N\geq2$, for which the orthogonal matrix $\mathcal{R}$ transforms nontrivially under the action of $G$, requiring a more careful analysis of boundary states.
We can categorize these boundary states based on both the structure of $\mathcal{R}$ and the values of $\vec{\phi}_D,\vec{\theta}_D$ into the following cases:
\begin{enumerate}
  \item Block-diagonal $\mathcal{R}=\mathcal{R}_1\oplus\mathcal{R}_2\oplus\cdots\oplus\mathcal{R}_k$ (with appropriate exchange of the components).
  \item Not block-diagonal and,
        \begin{enumerate}
          \item for all $a\in G$, $a\vec{\phi}_D\neq\vec{\phi}_D$ and $a\vec{\theta}_D\neq\vec{\theta}_D$,
          \item $\vec{\phi}_D=-\vec{\phi}_D$ and $\vec{\theta}_D=-\vec{\theta}_D$ under the compactification.
        \end{enumerate}
\end{enumerate}
Case 1 corresponds to the situation where the action of $G$ is disconnected, and the boundary states are factorized into boundary states of the connected subspace of $G$.
Case 2(a) corresponds to the boundary states with generic zero modes, which can be expressed as
\begin{equation}\label{case1orb}
  \ket*{\mathcal{R}(\vec{\phi}_D,\vec{\theta}_D)}_{\mathrm{orb}} = \frac{1}{\sqrt{\abs{G}}}\sum_{a\in G}D(a)\ket*{\mathcal{R}(\vec{\phi}_D,\vec{\theta}_D)}_c.
\end{equation}
In contrast, case 2(b) corresponds to the zero-mode parameters at the fixed points $\vec{\phi}_D=\vec{\phi}_E\in\pi\Lambda\cap\mathcal{V}_D$ and $\vec{\theta}_D=\vec{\theta}_E\in\pi\Lambda^\ast\cap\mathcal{V}_N$, for which the boundary states transform identically under both $a$ and $-a$, allowing us to make $G$-invariant states by symmetrizing over the subgroup $G_0=G/\qty{\pm I}$.
In the similar manner as in the single-component case, one can construct the consistent boundary state in this case by including the boundary state in the twisted sector as follows:
\begin{equation}\label{eq:orbifold_boundary_fixed}
  \ket*{\mathcal{R}(\vec{\phi}_E,\vec{\theta}_E)}_{\mathrm{orb}}
  =\sum_{b\in G_0}D(b)\qty[\frac{1}{\sqrt{\abs{G}}}\ket*{\mathcal{R}(\vec{\phi}_E,\vec{\theta}_E)}_c \pm 2^{-N/4}\ket*{\mathcal{R}(\vec{\phi}_E,\vec{\theta}_E)}_t],
\end{equation}
where the coefficients are determined from the consistency conditions.

Building on these general constructions, we can now determine the boundary states $\ket*{\Gamma_{1,2}}_{\mathrm{orb}}$, which are necessary to analyze the SRE.
First, we recall that $\ket*{\Gamma_1}_{\mathrm{orb}}$ corresponding to the mixed boundary condition in Eq.~\eqref{eq:gamma1_description} has the zero modes $\vec{\phi}_D=\vec{\phi}_E=\vec{0}$ and $\vec{\theta}_D=\vec{\theta}_E=\vec{0}$, which are at the fixed points.
We can thus use Eq.~\eqref{eq:orbifold_boundary_fixed} to obtain
\begin{equation}\label{eq:boundary_state_gamma_1_orb}
  \ket*{\Gamma_1}_{\mathrm{orb}}
  = \frac{1}{\sqrt{\abs{G}}}\sum_{b\in G_0}D(b)\ket*{\Gamma_1}_{c} \pm 2^{-N/4}\sum_{b\in G_0}D(b)\ket*{\Gamma_1}_t,
\end{equation}
where $\ket*{\Gamma_1}_t$ represents the corresponding boundary state in the twisted sector.
The boundary state $\ket*{\Gamma_2}_{\mathrm{orb}}$ in Eq.~\eqref{eq:gamma2_description} corresponding to the artificially created boundary is factorized into two-component boundary states, each of which is equivalent to $\ket*{\Gamma_1}_{\mathrm{orb}}$ at $\alpha=1$.

For later use, we also explicitly construct the twisted sector boundary states.
The boson field $\vec{\phi}$ in the twisted sector can be expanded as follows:
\begin{equation}
  \vec{\phi}(x,t)
  =\vec{\phi}_0 + \sum_{\substack{r\in\mathbb{Z}{+}\frac{1}{2} \\ r>0}}\frac{1}{\sqrt{4r}}\qty(\vec{b}_r^L e^{-ik_r(x+t)} + \vec{b}_r^R e^{ik_r(x-t)} + \text{H.c.}),
\end{equation}
with oscillator modes satisfying $[(\vec{b}_r^s)_i,(\vec{b}_u^t)_j]=\delta_{st}\delta_{ru}\delta_{ij}$.
Here $\vec{\phi}_0$ takes values in $\pi\Lambda_0$, where $\Lambda_0$ is the unit-cell of $\Lambda$:
\begin{equation}
  \Lambda_0=\qty{\vec{x}\;|\;\vec{x}=\sum_{i=1}^{N}\varepsilon_i\vec{e}_i,\;\varepsilon_i=0,1}.
\end{equation}
The Hamiltonian in the twisted sector is:
\begin{equation}
  H_t = \frac{2\pi}{L}\qty(\sum_{\substack{r\in\mathbb{Z}{+}\frac{1}{2}\\ r>0}} r[(\vec{b}_r^L)^\dag\cdot\vec{b}_r^L + (\vec{b}_r^R)^\dag\cdot\vec{b}_r^R] + \frac{N}{24}).
\end{equation}
The oscillator vacua are labeled either by the eigenvalues of $\vec{\phi}_0$ as $\ket*{\pi\vec{R}}_t\,(\vec{R}\in\Lambda_0)$ or by its dual $\vec{\theta}_0$ as $\ket*{\pi\vec{K}}_t\,(\vec{K}\in\Lambda^\ast_0)$, where $\Lambda_0^\ast$ is defined similarly to $\Lambda_0$.
These vacua are related by:
\begin{equation}
  \ket*{\pi\vec{R}}_t  = \frac{1}{\sqrt{2^N}}\sum_{\vec{K}\in\Lambda^\ast_0}e^{-i\pi\vec{K}\cdot\vec{R}}\ket*{\pi\vec{K}}_t,\quad
  \ket*{\pi\vec{K}}_t  = \frac{1}{\sqrt{2^N}}\sum_{\vec{R}\in\Lambda_0}e^{i\pi\vec{K}\cdot\vec{R}}\ket*{\pi\vec{R}}_t.
\end{equation}
The boundary states in the untwisted sector are eigenstates of the zero-mode operators $\mathcal{P}_D\vec{\phi}_0,\mathcal{P}_N\vec{\theta}_0$, and we need to obtain the corresponding boundary states also in the twisted sector.
Since the compactification lattice $\Lambda$ is a square lattice, the state $\ket*{\pi\vec{R}}_t$ is in the form of a tensor product of the boundary states for each component: $\ket*{\pi\vec{R}}_t=\ket*{\pi R_1\varepsilon_1}\ket*{\pi R_2\varepsilon_2}_t\cdots\ket*{\pi R_N\varepsilon_N}_t$.
We can exchange the basis of this Hilbert space so that the state $\ket*{\pi\vec{R}}_t$ is expressed as a tensor product $\ket*{\pi\mathcal{P}_D\vec{R}}_t\ket*{\pi\mathcal{P}_N\vec{R}}_t$.
Then, we rotate the state in $\mathcal{V}_N$ to create the eigenstate of the operator $\mathcal{P}_N\vec{\theta}_D$ as
\begin{equation}
  \ket*{\pi\mathcal{P}_N\vec{K}}_t = \frac{1}{\sqrt{2^{\mathrm{dim}\mathcal{V}_N}}}\sum_{\vec{R}\in\mathcal{P}_N\Lambda_0}e^{i\pi\vec{K}\cdot\vec{R}}\ket*{\pi\mathcal{P}_N\vec{R}}_t.
\end{equation}
The boundary state $\ket*{\pi\mathcal{P}_D\vec{R}}_t\ket*{\pi\mathcal{P}_N\vec{K}}_t$ is now an eigenstate of the two operators $\mathcal{P}_D\vec{\phi}_0,\mathcal{P}_N\vec{\theta}_0$ by construction.
Then, the boundary state $\ket*{\mathcal{R}(\vec{\phi}_E,\vec{\theta}_E)}_t$ in the twisted sector corresponding to the fixed point boundary state $\ket*{\mathcal{R}(\vec{\phi}_E,\vec{\theta}_E)}$ ($\vec{\phi}_E\in\pi\Lambda_0\cap\mathcal{V}_E,\vec{\theta}_E\in\pi\Lambda_0^\ast\cap\mathcal{V}_N$) in the untwisted sector is expressed as
\begin{equation}
  \ket*{\mathcal{R}(\vec{\phi}_E,\vec{\theta}_E)}_t = S_t(\mathcal{R})\ket*{\vec{\phi}_E}_t\ket*{\vec{\theta}_E}_t,
\end{equation}
where
\begin{equation}
  S_t(\mathcal{R}) = \exp(-\sum_{r}(\vec{b}_r^L)^\dag\cdot\mathcal{R}(\vec{b}_r^R)^\dag)
\end{equation}
is the squeezing operator in the twisted sector.
The group $G$ acts on the twisted sector in the same way as it acts on the untwisted sector boundary states: $D(a)\ket*{\mathcal{R}(\vec{\phi}_E,\vec{\theta}_E)}_t = \ket*{a\mathcal{R}a(a\vec{\phi}_E,a\vec{\theta}_E)}_t$.

\subsection{Calculation of the scaling dimensions}
The scaling dimension of a BCCO can be extracted from the transition amplitude between boundary states.
It equals the lowest conformal weight appearing in the decomposition of the amplitude into Virasoro characters.

For the pairs of Cardy states in the Ising CFT, we can express their boundary states in terms of the $S^1/\mathbb{Z}_2$ free-boson CFT boundary states as follows:
\begin{align}
  \ket*{ff}                   & = \ket*{D(\pi/2)}_{\mathrm{orb}} = \frac{1}{\sqrt{2}}\qty(\ket*{D(\pi/2)}_c + \ket*{D(-\pi/2)}_c), \\
  \ket*{\uparrow\uparrow}     & = \ket*{D(0)+}_{\mathrm{orb}} = \frac{1}{\sqrt{2}}\ket*{D(0)}_c + 2^{-1/4}\ket*{D(0)}_t,           \\
  \ket*{\downarrow\downarrow} & = \ket*{D(0)-}_{\mathrm{orb}} = \frac{1}{\sqrt{2}}\ket*{D(0)}_c - 2^{-1/4}\ket*{D(0)}_t.
\end{align}
Since the boundary states in the partition function $Z_{2\alpha}$ (Eq.~\eqref{eq:sre-partition-function}) are a $2\alpha$-fold product of these pairs, they can be written as
\begin{align}
  \ket*{\Gamma_f}          & = \ket*{D(\pi/2)}_{\mathrm{orb}}^{\otimes N} = \frac{1}{\sqrt{\abs{G}}}\sum_{a\in G}D(a)\ket*{D((\pi/2)\vec{d})}_c,             \\
  \ket*{\Gamma_\uparrow}   & = \ket*{D(0)+}_{\mathrm{orb}}^{\otimes N} = \frac{1}{\sqrt{\abs{G}}}\ket*{D(\vec{0})}_c + \cdots + 2^{-N/4}\ket*{D(\vec{0})}_t, \\
  \ket*{\Gamma_\downarrow} & = \ket*{D(0)-}_{\mathrm{orb}}^{\otimes N} = \frac{1}{\sqrt{\abs{G}}}\ket*{D(\vec{0})}_c + \cdots + 2^{-N/4}\ket*{D(\vec{0})}_t.
\end{align}

Let us first calculate the scaling dimension $h_{1f}$, which can be extracted from the amplitude between $\ket*{\Gamma_1}$ and $\ket*{\Gamma_f}$:
\begin{align}
  Z_{1f}(q)
   & = \frac{1}{\abs{G}}\sum_{a\in G}\sum_{b\in G_0}{}_c\!\mel*{D((\pi/2)\vec{d})}{D(a)e^{-\frac{\beta}{2}H} D(b)}{\Gamma_1}_c\notag                                                                                                                                    \\
   & = \frac{1}{2}\sum_{a\in G} {}_c\!\mel*{D((\pi/2)a\vec{d})}{e^{-\frac{\beta}{2}H}}{\Gamma_1}_c\notag                                                                                                                                                                \\
   & = \frac{1}{2}\sum_{a\in G}\frac{g_D g_1^{\mathrm{circ}}}{(\eta(q))^{N-1}}\sqrt{\frac{\eta(q)}{\theta_2(q)}} \sum_{\vec{K}\in\Lambda_\mathcal{R}^\ast} e^{i\vec{K}\cdot\mathcal{P}_D\vec{\phi}_D} q^{\vec{K}^2/8}\quad (\vec{\phi}_D = \frac{\pi}{2}a\vec{d})\notag \\
   & = \frac{1}{2}\sum_{a\in G}\frac{1}{(\eta(\tilde{q}))^N}\qty(\sum_{n=1}^{\infty}\tilde{q}^{\frac{1}{4}\qty(n-\frac{1}{2})^2})\sum_{\vec{R}\in\tilde{\Lambda}_{\mathcal{R}}} \tilde{q}^{2\qty(\vec{R}-\frac{1}{4}\mathcal{P}_D a\vec{d})^2}.
\end{align}
The lowest conformal weight $h=1/16$ comes from the term with $a=\pm I$ in the sum, which leads to
\begin{equation}
  h_{1f} = \frac{1}{16}.
\end{equation}

Similarly, we can also explicitly calculate the amplitude between $\ket*{\Gamma_1}$ and $\ket*{\Gamma_{\uparrow/\downarrow}}$.
Since $N=2\alpha$ is even, the sign in front of the twisted sector boundary state does not matter, and both $\uparrow$ and $\downarrow$ yield the same amplitude.
The calculation reads
\begin{align}\label{eq:amplitude-1up}
  Z_{1\uparrow/\downarrow}(q)
   & = \sum_{b\in G_0}\qty[\frac{1}{\abs{G}}\, {}_c\!\mel*{D(\vec{0})}{e^{-\frac{\beta}{2}H} D(b)}{\Gamma_1}_c + 2^{-N/2}\,{}_t\!\mel*{D(\vec{0})}{e^{-\frac{\beta}{2}H_t}D(b)}{\Gamma_1}_t]\notag \\
   & = \frac{1}{2}\frac{1}{(\eta(\tilde{q}))^N}\qty(\sum_{n=1}^{\infty}\tilde{q}^{\frac{1}{4}\qty(n-\frac{1}{2})^2})\sum_{\vec{R}\in\tilde{\Lambda}_{\mathcal{R}}}\tilde{q}^{2\vec{R}^2}
  + \frac{1}{2}\frac{\tilde{q}^{1/16}}{(\eta(\tilde{q}))^N}\qty(\sum_{n\in\mathbb{Z}}(-1)^n\tilde{q}^{n^2+n/2})\qty(\sum_{n\in\mathbb{Z}}(-1)^n\tilde{q}^{n^2})^{N-1}.
\end{align}
Here, the contribution from the twisted sector is computed as
\begin{align}
  {}_t\!\mel*{D(\vec{0})}{e^{-\frac{\beta}{2}H_t}}{\Gamma_1}_t
   & = q^{N/48}\qty(\frac{1}{\sqrt{2}}\prod_{n=1}^{\infty}\frac{1}{1+q^{n-1/2}})\qty(\prod_{n=1}^{\infty}\frac{1}{1-q^{n-1/2}})^{N-1}\notag                            \\
   & = \frac{\theta_2(e^{i\pi/2},q^{1/2})}{2\eta(q)}\qty(\frac{\theta_2(q^{1/2})}{2\eta(q)})^{N-1}\notag                                                               \\
   & = \frac{\tilde{q}^{1/16}}{\sqrt{2}\eta(\tilde{q})}\theta_4(\tilde{q}^{1/2},\tilde{q}^2)\qty(\frac{\theta_4(\tilde{q}^2)}{\sqrt{2}\eta(\tilde{q})})^{N-1}\notag    \\
   & = 2^{-N/2}\frac{\tilde{q}^{1/16}}{(\eta(\tilde{q}))^N}\qty(\sum_{n\in\mathbb{Z}}(-1)^n\tilde{q}^{n^2+n/2})\qty(\sum_{n\in\mathbb{Z}}(-1)^n\tilde{q}^{n^2})^{N-1},
\end{align}
where the theta functions $\theta_2(y,q)$ and $\theta_4(y,q)$ are defined as follows:
\begin{align}
  \theta_2(y,q) & = \sum_{n\in\mathbb{Z}} y^{n+1/2} q^{\frac{1}{2}\qty(n-\frac{1}{2})^2}, \\
  \theta_4(y,q) & = \sum_{n\in\mathbb{Z}} (-1)^n y^n q^{n^2/2}.
\end{align}
These two functions are related by the modular transformation as
\begin{equation}
  \theta_2(y,q) = \frac{1}{\sqrt{-i\tau}} \theta_4(y^{-1/\tau},\tilde{q}) e^{-\pi i z^2/\tau} \quad (y = e^{2\pi iz}).
\end{equation}
The lowest conformal weight contained in the amplitude~\eqref{eq:amplitude-1up} is $h=1/16$, and therefore the scaling dimensions are
\begin{equation}
  h_{1\uparrow} = h_{1\downarrow} = \frac{1}{16}.
\end{equation}

We can also calculate the scaling dimensions of the BCCO between $\ket*{\Gamma_2}$ and $\ket*{\Gamma_{f,\uparrow,\downarrow}}$ and confirm that the coefficient of the logarithmic contribution in the partition function is indeed zero, as expected from the fact that $\ket*{\Gamma_2}$ is the artificial boundary created by the folding.
To see this, the key point is that the scaling dimensions $h_{1f}=h_{1\uparrow}=h_{1\downarrow}$ do not depend on the number of components $N=2\alpha$.
Since $\ket*{\Gamma_2}$ is the $\alpha$-fold product of $\ket*{\Gamma_1}$ at $\alpha=1$, we can deduce that the scaling dimensions also multiply by $\alpha$.
Thus, we have
\begin{equation}
  h_{2f} = h_{2\uparrow} = h_{2\downarrow} = \frac{\alpha}{16},
\end{equation}
and the coefficient reads
\begin{equation}
  \gamma^{2f} = \gamma^{2\uparrow} = \gamma^{2\downarrow} = \frac{2\alpha}{24}\qty(\frac{1}{2} - 2) + 2\times \frac{\alpha}{16} = 0.
\end{equation}

We note that if one can identify the boundary state in the two-component $S^1/\mathbb{Z}_2$ free-boson CFT that describes the boundary obtained by folding the product of topological defects $1\otimes 1,\eta\otimes \eta,\mathcal{D}\otimes \mathcal{D}$ in the two-component Ising CFT, it should be possible to confirm that the conical singularity is absent even when we fold at the topological defect to make the geometry a rectangle.
To the best of our knowledge, in contrast to the single-component Ising CFT~\cite{oshikawa1997boundary}, such correspondence is not established, and we leave this problem for future work.

\section{\label{sec:defect_identification_fusion}Using SRE for a Systematic Identification of Lattice Realizations and Fusion Rules}

In this section, we establish a systematic procedure to identify lattice realizations of topological defects and determine their fusion rules without prior knowledge, utilizing the SRE as a diagnostic probe.
This approach allows for the discovery of non-invertible symmetries and their algebraic structures directly from the lattice Hamiltonian.

We search for topological defects in the critical Ising chain by considering a local modification of the Hamiltonian, specifically on the bond connecting sites $L$ and $1$.
We assume the defect Hamiltonian $H_{L1}$ acts non-trivially only on these two sites and takes a general form of the sum of Pauli strings:
\begin{equation}
  H_{L1} = e_{1}\sigma^{i}\otimes I+e_{2}I\otimes\sigma^{j}+e_{3}\sigma^{k}\otimes\sigma^{l},
\end{equation}
where the coefficients take values $e_1,e_2 \in \{0,\pm 1\}$ and $e_3 \in \{\pm 1\}$, and the operators are chosen from $\sigma^{i,j,k,l} \in \{X,Y,Z\}$.
These constraints on the coefficients are physically motivated by the properties of the Clifford group.
Since topological defects can typically be moved or fused via Clifford unitaries, and since Clifford unitaries map (Hermitian) Pauli strings to (Hermitian) Pauli strings with coefficients $\pm 1$, it is natural to consider the lattice realizations of such defects to be composed of terms with integer coefficients.
Furthermore, to ensure the defect represents a valid interface within the chain, we impose the condition that $H_{L1}$ does not decouple the chain or trivially block-diagonalize the Hamiltonian; for instance, a term like $X_L Z_1$ is excluded as it would isolate the operator $Z_1$ from the interaction.

Based on these criteria, we identify $648$ distinct candidate defect configurations.
To efficiently analyze these candidates, we classify them into equivalence classes based on Clifford unitaries.
Specifically, we consider the local Clifford group $G = \langle S_L, S_1, \text{CZ}_{L,1} \rangle$ that preserves the form of the neighboring bulk interaction terms ($Z_{L-1}Z_L$ and $Z_1 Z_2$) by mapping the boundary $Z$ operators to $\pm Z$.
Here, $S_L,S_1$ are the phase gates acting on sites $L$ and $1$, respectively, and $\text{CZ}_{L,1}$ is the controlled-$Z$ gate between these two sites.
This classification reduces the vast search space of $648$ candidates to $44$ distinct equivalence classes, which include the identity defect($-X_L-X_1-Z_LZ_1$), the $\eta$ defect ($-X_L-X_1+Z_LZ_1$), and the duality defect ($-X_L-Z_1 - X_1 Z_L$).

We then determine the topological nature of each candidate by analyzing the finite-size scaling of the SRE, $M_\alpha(L)$.
The SRE serves as a sensitive probe for the properties of the defect.
If the defect is topological, the SRE must exhibit a universal size-independent subleading term, $c_\alpha^{\mathcal{A}}$, governed by the $g$-factor of the boundary state $\Gamma_1$ in the $\mathcal{A}$ defect sector.
This is manifested as a clear data collapse of the constant term across different system sizes near the critical point.
Conversely, if the defect corresponds to or flows to a non-topological defect, the geometry of the replicated partition function after the folding acquires sharp corners.
This results in a universal logarithmic correction to the SRE.

\begin{figure}[tb]
  \centering
  \includegraphics[width=0.48\linewidth]{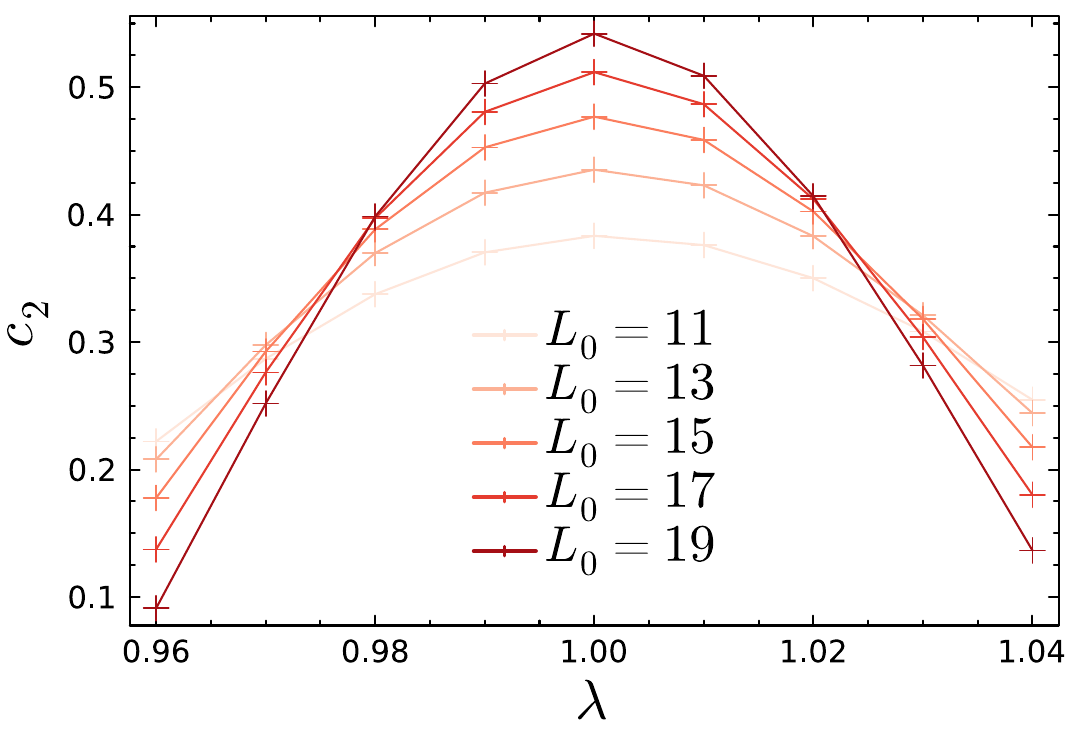} \hfill
  \includegraphics[width=0.48\linewidth]{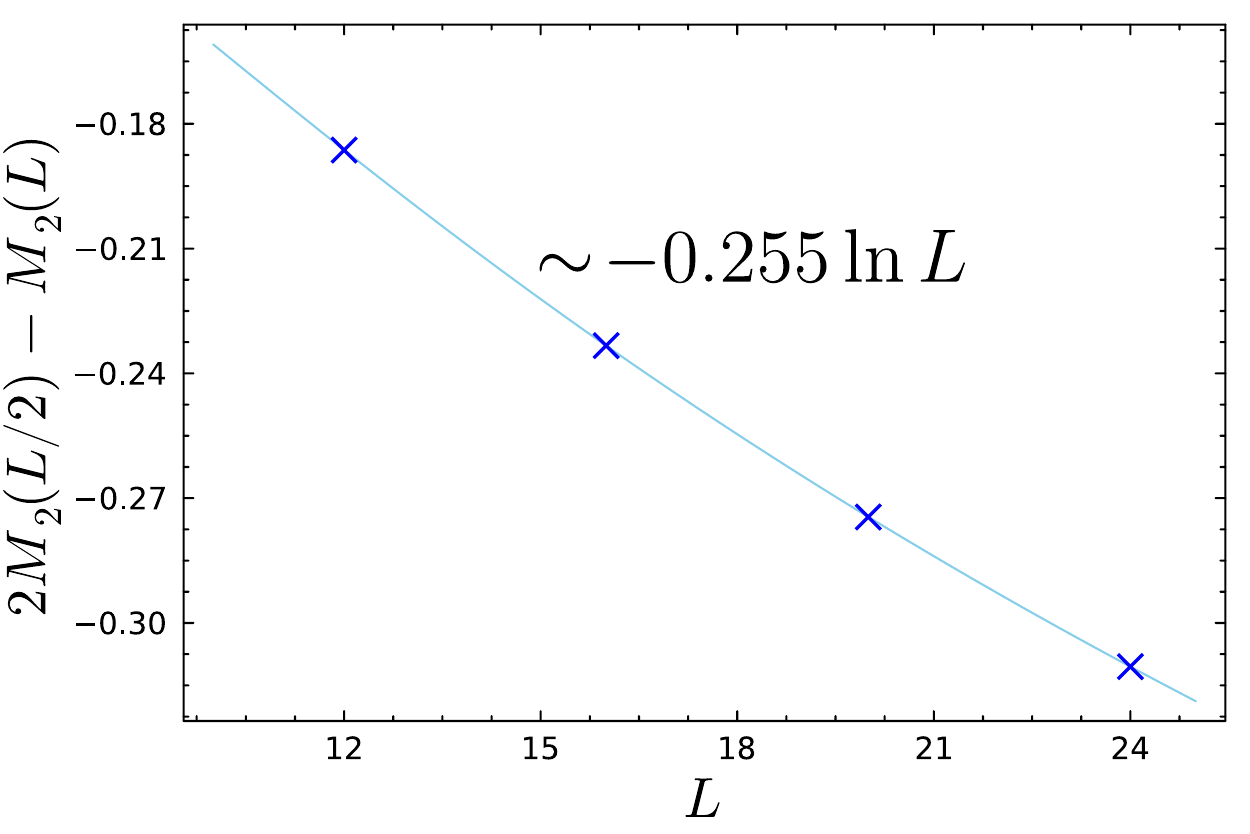}
  \caption{(Left) Finite-size scaling of the constant term $c_2$ extracted from the SRE of the Ising model with the candidate defect $H_{L1} = -Z_L - X_1 - X_L Z_1$. The lack of data collapse at the critical point signals that the defect is not topological. (Right) Extraction of the logarithmic term from the difference $2M_2(L/2) - M_2(L)$. The estimated coefficient $\approx -0.255$ is close to the value $-1/4$ characteristic of factorizing defects (open boundaries).}
  \label{fig:c2_nontopo}
  \label{fig:log_nontopo}
\end{figure}

To demonstrate this approach, we analyze a specific candidate defect given by $H_{L1} = -Z_L - X_1 - X_L Z_1$, which does not correspond to any known topological defects.
We first test the hypothesis that the defect would be topological by fitting the $\alpha=2$ SRE data to the form $M_2(L) = m_2 L - c_2 + r/L$.
As shown in Fig.~\ref{fig:c2_nontopo}, the extracted constant $c_2$ shows significant finite-size dependence and fails to exhibit data collapse, indicating the defect is not topological. Subsequently, we test for the presence of a logarithmic contribution by computing the difference $2M_2(L/2) - M_2(L)$, which isolates the logarithmic term in leading order.
The results in Fig.~\ref{fig:log_nontopo} reveal a logarithmic dependence with a coefficient of approximately $-0.255$.
This value is close to the theoretical value of $-1/4$ expected for factorizing defects.
This suggests that the defect either flows to a factorizing defect in the IR limit, or is a non-topological, non-factorizing defect for which the BCCO involving $\Gamma_1$ has the same scaling dimension as that of the factorizing defects.

Once the lattice realizations of topological defects are systematically identified, their fusion rules can be determined by exploring the action of Clifford unitaries.
Let $\{D^1, D^2, \ldots, D^m\}$ be the set of identified topological defects.
Suppose that we have the adjacent defects $D_{L1}^i, D_{12}^j$.
To identify the fusion rule of these two defects, we exhaustively apply all possible Clifford unitaries to this defect term and calculate the resulting expression; note that the number of distinct Clifford unitaries acting on two sites is $11,520$ and on three sites is $92,901,120$, which is always finite.
We terminate this search process if either of the following results is obtained:
\begin{enumerate}
  \item The transformed term matches one of the lattice realizations of topological defects, say $D^k$. This outcome confirms the fusion rule $D^i \otimes D^j = D^k$.
  \item A Clifford unitary decouples a few sites, and the remaining terms match the expressions of topological defects $\qty{ D^{k_1}, D^{k_2}, \ldots, D^{k_l}}$. This outcome confirms the fusion rule $D^i \otimes D^j = D^{k_1} \oplus D^{k_2} \oplus \cdots \oplus D^{kl}$. For instance, the fusion rule ($\mathcal{D} \otimes \mathcal{D} = \mathcal{T}^{-}(1 \oplus \eta)$) corresponds to a Clifford unitary decoupling a single site $Z_1$ with the remaining terms being the identity ($-X_L-X_2-Z_LZ_2$) and the $\eta$ defect ($-X_L-X_2+Z_LZ_2$).
\end{enumerate}
We emphasize that, since the number of Clifford unitaries is finite and each calculation of the transformation is computationally inexpensive (due to the Gottesman-Knill theorem), this is a practical approach to finding the fusion rules. Importantly, this method of finding the fusion rules is guaranteed to terminate.
This procedure establishes a practical and rigorous method to discover the algebraic structure of generalized symmetries directly from lattice Hamiltonians.

\end{document}